\documentstyle[prl,aps,amsfonts,amsmath,preprint,floats,twoside]{revtex}

\input BoxedEPS
\SetRokickiEPSFSpecial  
\HideDisplacementBoxes
\tighten
\newcommand{\be}{\begin{equation}}
\newcommand{\bea}{\begin{eqnarray}}
\newcommand{\ba}{\begin{array}}
\newcommand{\ea}{\end{array}}
\newcommand{\en}{\end{equation}}
\newcommand{\eea}{\end{eqnarray}}
\newcommand{\half}{\fract{1}{2}}
\newcommand{\nf}{ v/\sqrt{N_{F}}}
\newcommand{\nff}{ v^2/N_F}
\newcommand{\dslash}{\partial \hskip -0.55em /}
\newcommand{\svec}[1]{\skew2\vec#1} 
\numberwithin{equation}{section} 
\newcommand{\fract}[2]{{\textstyle\frac{#1}{#2}}} 
\renewcommand{\thesection}{\arabic{section}} 
\begin{document}

\title{Heavy Fermion Stabilization of Solitons in  1+1
Dimensions}

\author{E.~Farhi\footnote[0]{e-mail:  farhi@mit.edu,
graham@pierre.mit.edu, jaffe@mit.edu,
weigel@ctp.mit.edu\vspace*{-1pc}}$^{\rm a}$, N.~Graham$^{\rm a,b}$,
R.L.~Jaffe$^{\rm a}$, and
H.~Weigel\footnote{Heisenberg Fellow}$^{\rm a}$}

\address{{~}\\$^{\rm a}$Center for Theoretical Physics, Laboratory for
	Nuclear Science,
	and Department of Physics \\
	Massachusetts Institute of Technology,
	Cambridge, Massachusetts 02139 \\
	and \\
	$^{\rm b}$Dragon Systems, Inc.,
	Newton, MA 02460\\
	{\rm MIT-CTP\#2950}}

\maketitle
\thispagestyle{empty} 

\begin{abstract}\noindent We find static solitons stabilized by
quantum corrections in a (1+1)-dimen\-sional model with a
scalar field chirally coupled to fermions.  This model does not
support classical solitons.  We compute the renormalized
energy functional including one-loop quantum corrections. 
We carry out a variational search for a configuration that
minimizes the energy functional.  We find a nontrivial
configuration with fermion number whose energy is
lower than the same number of free fermions quantized
about the translationally invariant vacuum.  In order to
compute the quantum corrections for a given background
field we use a phase-shift parameterization of the Casimir
energy.  We identify orders of the Born series for the phase
shift with perturbative Feynman diagrams in order to
renormalize the Casimir energy using perturbatively
determined counterterms.  Generalizing dimensional
regularization, we demonstrate that this procedure yields a
finite and unambiguous energy functional.
\end{abstract}


\section{Introduction}\label{sec:1}

Many quantum field theories contain solitons, nontrivial configurations
that are local minima of the energy functional.  Often, the existence of a
soliton can be anticipated from topological properties of the field theory.
In such cases, topologically nontrivial configurations cannot be
continuously deformed into a vacuum configuration.  Still, a detailed look
at the dynamics is required to see that such configurations do not collapse
to a point.  For example, the soliton in the Skyrme model is unstable
without the addition of a four-derivative term to the action.  In general, the
question of soliton stability requires a detailed study of the energy.  If
we have a reliable method for computing the leading quantum contributions
to the energy of nontrivial field configurations, we can begin to study the
stability of topological and nontopological solitons. The techniques
introduced in Ref.~\cite{method1,method2,method3,tbaglevi} provide such a
framework in renormalizable quantum field theories.

In this paper we apply these techniques to reexamine suggestions made in
previous works \cite{prevwork,Farhi} that a heavy fermion can create a
soliton. We consider a simple renormalizable quantum field theory in 1+1
dimensions.  In this model, a two-component scalar field couples to
fermions in a chirally invariant way.  The fermion mass arises from a
Yukawa coupling to the scalar, which has a nonzero vacuum expectation
value.  At the classical level the theory has no solitons.  (This is in
contrast to $N=2$ supersymmetric models considered in Ref.~\cite{vanN},
which support classical solitons that are unmodified by quantum
corrections.)  We investigate whether configurations of the scalar field
that carry fermion quantum numbers can be lighter than ordinary fermion
states built on top of the translationally invariant vacuum.  We show that
they can, and that the lightest of these configurations can be identified
as a stable soliton. 

It is well known that a fermion can be strongly bound in the background of
a spatially varying scalar field.  But the binding energy comes with the 
cost of the classical energy of the scalar configuration.  In addition, 
there is the possible cost of an increase of the fermion Casimir energy, 
since the fermion vacuum is polarized by the background scalar field.  We 
must include this contribution because it is of the same order in the 
loop-counting parameter as the binding energy of the strongly bound level.

We give a detailed description of our method for calculating the one-loop
fermion contribution to the energy of a fixed scalar background.  Our
method is exact for an arbitrary spatially varying background.  It sums all
orders in the derivative expansion.  Since we work in a renormalizable
theory, our results are unambiguous; we use the same counterterms in the
presence of the soliton that we determined from the renormalization
conditions in the perturbative sector.  In order to combine the infinite
loop energy and the infinite counterterm contribution into a finite answer,
we must regularize the theory.  Phase shifts are a key ingredient in our
approach.  The Casimir energy is the sum over the energy shift of each mode
in the fixed background.  We express the sum over modes as an integral over
the continuum density of states, which we relate in turn to the phase
shifts.  Then the ultraviolet-divergent pieces of the loop energy are
linked to low orders of the Born series for the phase shifts, which can be
unambiguously identified with Feynman diagrams.  In this framework the
quantum corrections to the energy are expressed as the sum of two finite
pieces: a momentum integral involving Born-subtracted phase shifts, and a
small number of perturbatively renormalized Feynman diagrams, which
correspond to the Born subtractions~\cite{method1,method2,method3}.

The (1+1)-dimensional model we consider has many features in common with
the linear $\sigma$-model in 3+1 dimensions: renormalizability, chirally
symmetric scalar-fermion interactions, the possibility of a Dirac
Hamiltonian with a spectrum that is asymmetric in energy,
and a topological structure. Hence our findings in the (1+1)-dimensional
case suggest that similar phenomena might occur in $3+1$ dimensions: heavy
fermions can be realized as soliton configurations carrying fermion
number.  This result is in agreement with decoupling theorems \cite{Farhi},
which show that solitons supply the necessary fermion numbers to maintain
anomaly cancellation after a fermion is decoupled by increasing its Yukawa
coupling.

In a theory with $N_F$ fermions, quantum corrections due to boson loops are
suppressed by a factor of $N_F$ relative to the fermion quantum
corrections.  Therefore we work with  $N_F$ large. However, boson loops
require special attention in a (1+1)-dimensional model with (classical)
spontaneous symmetry breaking.  Because of infrared singularities, the
symmetry is actually restored by these corrections and it would be
inappropriate to construct a soliton about a classical minimum.  To
circumvent this difficulty we introduce an explicit symmetry breaking term
that is large enough to guarantee a unique vacuum state.  Other treatments
of solitons at the quantum level based on the inverse scattering method
\cite{Shei,Zee,Campbell} cannot accommodate this essential ingredient.

Our paper is organized as follows: In Section~\ref{sec:2} we introduce the scalar
sector of the model and explain the necessity of explicitly breaking the
chiral symmetry.  Section~\ref{sec:3} contains the discussion of the fermionic
sector. In Section~\ref{sec:4} we combine the scalar and fermion sectors to
form the full expression for the energy.  The corresponding numerical results
are presented in Section~\ref{sec:5}.  We summarize our findings in
Section~\ref{sec:6}.  In Appendix~\ref{app:a} we justify the phase shift method
using dimensional regularization.  In Appendix~\ref{app:b} we demonstrate
the behavior of the strongly bound fermion level in the WKB approximation. 
In Appendix~\ref{app:c} we explain why the soliton background need not be
reflectionless, as claimed in previous works.  Some of the results presented in
this paper were summarized in Ref.~\cite{letter}.

\section{The Boson Sector}\label{sec:2}

We consider a two component scalar field governed by the Lagrangian
\begin{equation}
{\cal L}_B=\fract{1}{2}\, \partial_\mu\svec\phi\cdot\partial^\mu\svec\phi
-V(\svec\phi)
\label{lagb}
\end{equation}
where $\svec\phi=(\phi_1,\phi_2)$.  The potential $V$
contains the usual Higgs term and an
explicit symmetry breaking piece proportional to $\phi_1$,
\begin{equation}
V(\svec\phi)=\frac{\lambda}{8}
\Bigl(\svec\phi\cdot\svec\phi
- v^2+\frac{2\alpha v^2}{\lambda}\Bigr)^2
-\frac{\lambda}{2}\Bigl(\frac{\alpha v^2}{\lambda}\Bigr)^2
-\alpha v^3 (\phi_1- v)\, .
\label{lag2}
\end{equation}
If $\alpha=0$, the $U(1)$ transformation
\begin{equation}
\phi_1+i\phi_2 \longrightarrow e^{i\varphi}
\left(\phi_1+i\phi_2\right)
\label{u1sym}
\end{equation}
is a symmetry of ${\cal L}_{B}$, but we take $\alpha$ big enough to
avoid infrared singularities that would arise when this symmetry is
spontaneously broken.  For $\alpha>0$, $V(\svec\phi)$ has a unique
classical minimum at $\svec\phi_{\rm cl}=(v,0)$. The modes fluctuating
about the classical vacuum have masses
\begin{equation}
m_\sigma^2=\left(\lambda+\alpha\right) v^2
\quad {\rm and} \quad
m_\pi^2=\alpha v^2 \, .
\label{masses}
\end{equation}
Note that on the chiral circle, $\svec\phi\cdot\svec\phi=
v^2$, this model reduces to the sine-Gordon model.

Although we are principally interested in the stabilizing effects of
fermions, we must first consider the infrared singularities in the boson
sector that could invalidate our approach.  We wish to study scalar field
configurations that are deformations of the classical vacuum.  However, it
is well known that when $\alpha=0$, spontaneous symmetry breaking does not
occur~\cite{Coleman1d}.  The absence of long-range order in 1+1 dimensions
can be traced to infrared singularities associated with the propagator of
the massless mode.  Making $\alpha$ big enough gives a large enough mass to
the would-be Goldstone mode to stabilize the classical vacuum.

The infrared problems can be seen in the renormalized one-loop effective
potential, $V^{B}_{\rm eff}$,
\begin{equation}
V^{B}_{\rm eff}(\svec\phi)=
\frac{1}{16\pi}\biggl\{
\kappa_+^2\Bigl[1-\ln\Bigl(\frac{\kappa_+^2}{m_\sigma^2}\Bigr)\Bigr]
+\kappa_-^2\Bigl[1-\ln\Bigl(\frac{\kappa_-^2}{m_\pi^2}\Bigr)\Bigr]
-m_\sigma^2-m_\pi^2\biggr\} 
\label{bveff}
\end{equation}
where $\kappa_\pm^2$ are the eigenvalues of the $2\times2$ matrix
$\partial^2 V/\partial \phi_i\partial \phi_j$:
\begin{equation}
\kappa_{\pm}^2(\svec\phi\,^2)
=\frac{\lambda}{2}\bigl(2\svec\phi\,^2\pm\svec\phi\,^2- v^2
+2\alpha v^2/\lambda\bigr)\, .
\label{meigval}
\end{equation} We have renormalized $V^{B}_{\rm eff}$ so that the vacuum
expectation value at $\svec\phi=\svec\phi_{\rm cl}$ and the particle
masses $m_{\sigma}$ and $m_{\pi}$ remain unchanged.  As $\alpha\to 0$,
$m_{\pi}^{2}\to 0$, so $V^{B}_{\rm eff}$ develops a logarithmic
infrared divergence reflecting the fluctuations that restore $U(1)$
invariance.  But for any nonzero $\alpha$, $V^{B}_{\rm eff}$ is
well defined and has a unique minimum at $\svec\phi_{\rm cl}$.  

When we consider the fermion one-loop contribution, we find that when
$\alpha$ becomes small the character of configurations that minimize the
energy changes. To see this, note that $V^{B}_{\rm eff}(\svec\phi)=0$ for
any scalar field configuration on the chiral circle.  As $\alpha\to 0$,
$\svec\phi$ is forced to the chiral circle because any other choice has
infinite energy.  On the other hand, for a slowly varying $\svec\phi$ that
circles zero once as $x$ goes from $-\infty$ to $+\infty$, the classical
energy goes to zero if $\svec\phi$ stays on the chiral circle and the scale
over which $\svec\phi$ varies gets large.  This nontrivial field
configuration has zero boson contribution to its energy through one loop in
the limit $\alpha\to 0$.  (Since the configuration varies slowly,
integrating eq.~(\ref{bveff}) over $x$ gives a good approximation to the
one-loop effective energy; because we are in 1+1~dimensions, all higher
derivative terms go to zero as the scale over which $\svec\phi$ varies gets
large.)  When we consider the fermion one-loop contribution to the energy
of this configuration, we find that it too contributes zero.  Furthermore this
configuration has fermion number one.  Thus as
$\alpha\to 0$ the model has a zero-energy, infinitely large soliton
carrying fermion number.  We believe this object is destroyed by the
fluctuations that restore $U(1)$ invariance.  In our numerical search for
solitons, we find a sharp transition as we vary $\alpha$.  At small
$\alpha$, we find that configurations with large widths are energetically
favored, and the width is controlled by $\alpha$. At large $\alpha$ we find
configurations with moderate widths are favored, and the width is no
longer sensitive to changes in $\alpha$.  We are interested in the latter.

\section{The Fermion Sector}\label{sec:3}

The purpose of this section is fourfold.  First, we set up the formal
expression for the contribution of the fermion sector to the energy, which
is expressed through the eigenvalues of the single-particle Dirac equation.
Second, we set up the formalism for solving the Dirac equation in a general
background $\svec\phi(x)$.  We show how to obtain the phase shifts, which
characterize the continuum states.  Third, we explain how the properly
regulated and renormalized fermion one-loop contribution to the energy can
be computed in terms of the phase shifts and the bound state energies.  And
fourth, we explain how the fermion number of the background $\svec\phi\,$
can be computed in terms of the phase shifts and bound states.

\subsection{The Fermion Casimir Energy}

We introduce $N_F$ Dirac fermions coupled in a chirally invariant way
to $\svec\phi$. We suppress the flavor label, but keep track of factors
of $N_{F}$ as necessary,
\begin{equation}
{\cal L}_F=\frac{i}{2}\left[\bar{\Psi},\dslash\Psi\right]
-\frac{G}{2}\left(\left[\bar{\Psi},\Psi\right]\phi_1
+i\left[\bar{\Psi},\gamma_5\Psi\right]\phi_2\right)
\label{lagf}
\end{equation}
with $\bar\Psi = \Psi^\dagger \gamma^0$. The fermions
acquire mass $m=G v$ when the boson field takes the vacuum value
$\svec\phi_{\rm cl}$.  We have been careful about operator ordering in
eq.~(\ref{lagf}) because of its role in determining the fermionic
contribution to the vacuum energy.  The ordering of anticommuting fermion
fields is fixed by requiring charge conjugation invariance.  We define
${\cal C}\phi_{1} = \phi_{1}$, ${\cal C}\phi_{2} = -\phi_{2}$, and choose a
Majorana basis for the Dirac matrices, $\gamma^0= \sigma_2$,
$\gamma^1=i\sigma_3$, and $\gamma_5=\sigma_1$, so that ${\cal C}\Psi =
\Psi^{\ast}$.  With these definitions it is easy to verify that
$[\bar{\Psi},\Psi]$ is even under ${\cal C}$ and that
$[\bar{\Psi},\gamma_5\Psi]$ and $[\bar{\Psi},\gamma^{\mu}\Psi]$ are odd
under ${\cal C}$.

Given the ordering in eq.~(\ref{lagf}), the second quantized Hamiltonian
becomes ${\cal H}[\svec\phi\,]= \half\int dx$ $[\Psi^\dagger,
H(\svec\phi\,)\Psi]$, where $H(\svec\phi\,)$ is the single-particle
Dirac Hamiltonian in the presence of $\svec\phi(x)$. Denoting the
(positive and negative) energy eigenvalues of $H$ by $\omega_n$, and
the corresponding eigenfunctions by $\psi_n$, we make the usual Fock
decomposition, $\Psi(x,0)=\sum_{\omega_{n}>0} b_{n}\psi_{n}(x)+
\sum_{\omega_{n}<0}  d^{\,\dagger}_{n}\psi_{n}(x)$, so that
\begin{equation}
{\cal H}[\svec\phi\,] = 
\half\sum_{\omega_{n}>0}\omega_{n}(b^{\dagger}_{n}b_{n}
-b_{n}b^{\dagger}_{n}) +  \half\sum_{\omega_{n}<0}\omega_{n}(d_{n}
d^{\dagger}_{n}-d^{\dagger}_{n}d_{n}) \, .
\label{calH}
\end{equation}

We are interested in the expectation value of ${\cal H}[\svec\phi\,]$
in states of definite occupation number.  Consider the vacuum state
$|\Omega\rangle$, characterized by
$b_n|\Omega\rangle=d_n|\Omega\rangle=0$.  Then 
\begin{equation}
E^{F}[\svec\phi\,]=\langle \Omega | {\cal H}[\svec\phi\,]
|\Omega \rangle
=-\fract{1}{2}\sum_{n}|\omega_n| \, .
\label{energ_vac}
\end{equation}
Of course this sum is divergent.  Its regularization
and renormalization are discussed below.

It is easy to confirm that $|\Omega\rangle$ is the lowest energy state.  
We see that the vacuum energy in a charge conjugation invariant theory is
given by the (regularized) sum over the absolute values of \emph{all}
single-particle states. This result can be verified \cite{Herbert} using 
functional integral techniques, where it follows from the evaluation of the path
integral for large Euclidean times.  The fermion number of
$|\Omega\rangle$ depends on $\svec\phi$, and for many configurations of
interest it is not zero.

For completeness, we write out the one-loop fermion vacuum contribution 
to the effective potential,
\begin{equation}
V^{F}_{\rm eff}(\svec\phi)=
-\frac{G^2}{4\pi}\left\{\svec\phi\,^2
\left[1-\ln\bigl(\svec\phi^2/v^2\bigr)\right] -v^2\right\}\, .
\label{fveff}
\end{equation}
Of course, our principal motivation in this work is to go beyond
the effective potential approximation for the fermion one-loop
contribution to the energy.  Nevertheless, eq.~(\ref{fveff}) will be
useful to verify our numerical results for the fermion
one-loop vacuum contribution in the case of a slowly varying boson
fields.  Also, like the boson effective potential, $V^{F}_{\rm eff}$
vanishes on the chiral circle, as we anticipated in the discussion of
the $\alpha\to 0$ limit in the previous section.

\subsection{Solutions to the Dirac Equation in a Chiral Background}

The Dirac Hamiltonian is
\begin{equation}
H[\svec\phi\,] = i\sigma_1 \frac{d}{dx} + G\sigma_2\phi_1(x) +
G\sigma_3\phi_2(x).
\label{dirham}
\end{equation}
Although the underlying theory is charge conjugation invariant, the
Dirac Hamiltonian in the presence of a fixed $\svec\phi(x)$ is not.
Thus it is necessary to consider both positive and negative energy
eigenvalues $\omega$ of the time-independent Dirac equation
\begin{equation}
H \psi = \omega \psi \, .
\end{equation}
The associated second-order equations for the Dirac spinor,
$\psi \equiv \bigl(\begin{smallmatrix}
f\\ g
\end{smallmatrix})$, are
\begin{eqnarray}
-f'' - G\phi_1'f + G\phi_2'(\omega + G\phi_2)^{-1}(f' + G\phi_1f)&=&
(\omega^2 - G^2\phi_1^2 - G^2\phi_2^2)f\nonumber\\[1ex]
-g'' + G\phi_1'g - G\phi_2'(\omega - G\phi_2)^{-1}(g' - G\phi_1g)&=&
(\omega^2 - G^2\phi_1^2 - G^2\phi_2^2)g\, .
\label{deqfg}
\end{eqnarray}

In one dimension, there are two channels for each energy.  The
$S$-matrix is $2$ dimensional and, in general, not diagonal.
We simplify this situation using parity invariance.  By demanding that
$\phi_1$ and $\phi_2$ are respectively even and odd under coordinate
reflection, i.e., $\phi_1(x)=\phi_1(-x)$ and
$\phi_2(x)=-\phi_2(-x)$, we ensure that the Dirac Hamiltonian,
eq.~(\ref{dirham}), is invariant under parity,
\begin{equation}
[P,H] = 0,\quad {\rm where} \quad P= \gamma_0\Pi=\sigma_2\Pi 
\label{parity1}
\end{equation}
and $\Pi$ is the coordinate reflection operator that transforms $x$
to $-x$. Thus we can choose a basis of parity eigenstates,
\begin{equation}
P\psi_{\pm}(x) \equiv
\sigma_2\psi_{\pm}(-x) = \pm \psi_{\pm}(x)\ .
\label{parity2}
\end{equation}

Using parity we can replace the scattering problem on the line
$x\in [-\infty,\infty]$ by two scattering problems on the
half-line $x\in [0,\infty]$ corresponding to even and 
odd parity.  Eq.~(\ref{parity2}) gives boundary conditions 
at $x=0$ on the parity eigenstates:
\begin{equation}
 \psi_{+}(0) \propto \biggl(\begin{matrix}
1 \\ i\end{matrix}\biggr) \qquad {\rm and}\qquad
 \psi_{-}(0) \propto \biggl(\begin{matrix}
1 \\ -i
\end{matrix}\biggr)\, .
 \label{bc}
\end{equation}
The solution to the Dirac equation on the half-line with 
either boundary condition is unique up to an overall
normalization.  For $x\to\infty$, this unique solution can be written as a
superposition of incoming ($\propto e^{-ikx}$) and outgoing
($\propto e^{ikx}$) waves.  The coefficient of the outgoing wave 
relative to the incoming wave defines the phase shift.

To implement this program, we introduce eigenstates of the
free Dirac Hamiltonian with energy $\omega$,
\begin{eqnarray}
\varphi^{0}_{+k}(x) &=&\frac{1}{\omega}
\biggl(\begin{matrix}\omega\\ -k+im
\end{matrix}\biggr) e^{ikx}
\nonumber\\
\varphi^{0}_{-k}(x) &=&\frac{1}{\omega}
\biggl(\begin{matrix}\omega\\ k+im
\end{matrix}\biggr)
e^{-ikx} \,
\label{asymptotic}
\end{eqnarray}
where $k = +\sqrt{\omega^2 - m^2}$.  Next, we construct the
eigenstates of the full interacting Dirac Hamiltonian with energy $\omega$
that are asymptotic to $\varphi^{0}_{\pm k}$ as $x\to\infty$,
\begin{equation}
\varphi_{+k}(x)= \Biggl(\begin{matrix}f(x)\\
\displaystyle\frac{i}{\omega + G\phi_2(x)}(f'(x) + G\phi_1(x)f(x))
\end{matrix}\Biggr)
\label{fullsol2}
\end{equation}
and
\begin{equation}
\varphi_{-k}(x)= \Biggl(\begin{matrix}f^{\ast}(x)\\
\displaystyle\frac{i}{\omega + G\phi_2(x)}({f^{*}}'(x) + G\phi_1(x)f^{\ast}(x))
\end{matrix}\Biggr)
\label{fullsol1}
\end{equation}
where $f(x)$ is the solution to the real second-order equation,
eq.~(\ref{deqfg}), for the upper component, subject to the boundary
condition that $f(x) \to e^{ikx}$ as $x\to\infty$.  It is easy
to verify that in the same limit $\varphi_{\pm k}(x) \to
\varphi^{0}_{\pm k}(x)$ since the boson fields approach their vacuum
values.

Since the even and odd parity channels decouple, the $S$-matrix is
diagonal.  Its diagonal elements $S_\pm = e^{2i\delta_\pm(\omega)}$ can be
defined through the even and odd parity eigenstates of $H$,
\begin{eqnarray}
\psi_+(x) &=& \varphi_{-k}(x) + 
\frac{m-ik}{\omega}S_+(\omega)\varphi_{+k}(x)
\nonumber\\
\psi_-(x)&=&\varphi_{-k}(x) - 
\frac{m-ik}{\omega}S_-(\omega)\varphi_{+k}(x)\, .
\label{sdef}
\end{eqnarray}
If we set the interaction to zero ($\phi_1=v, \phi_2=0$), then $\psi_+$
($\psi_-$) reduces to the even (odd) parity solution to the free Dirac
equation with $S_\pm=1$, which explains the extra factor of
$\frac{m-ik}{\omega}$ in eq.~(\ref{sdef}).

To determine $S_{\pm}$ we use the fact that the eigenstates of eq.~(\ref{sdef})
obey eq.~(\ref{bc}). For the positive parity channel this yields
\begin{equation}
S_{+}(\omega) =
-\frac{(m+ik)\left[(\omega - G\phi_1(0))f^\ast(0) - {f^\prime}^\ast(0)\right]}
{\omega\bigl[(\omega - G\phi_1(0))f(0) - {f^\prime}(0)\bigr]} 
\label{splus1}
\end{equation}
and similarly for the negative parity channel
\begin{equation}
S_-(\omega) = 
\frac{(m+ik)\left[(\omega + G\phi_1(0))f^\ast(0) + {f^\prime}^\ast(0)\right]}
{\omega\bigl[(\omega + G\phi_1(0))f(0) + {f^\prime}(0)\bigr]}\, .
\label{sminus1}
\end{equation}

To compute the phase shifts efficiently and avoid $2\pi$ ambiguities,
it is convenient to factor the free solution out of $f(x)$ by writing
$f(x)=e^{ikx}e^{i\beta(x,\omega)}$.  Then the phase shifts are given~by
\begin{equation}
\delta_{\pm}(\omega) = -\mathop{\rm Re} \beta (0,\omega)-
\arg\Bigl[1+\frac{i\beta^\prime(0,\omega) + G(\phi_1(0) - v)}
{\mp\omega+m+ik}\Bigr] 
\label{beta}
\end{equation}
where the complex function $\beta (x,\omega)$ solves the differential
equation
\begin{eqnarray}
-i\beta''(x,\omega)+2k\beta'(x,\omega)+\beta^{\prime2}(x,\omega) - m^2 +
G^2\phi_1^2(x) + G^2\phi_2^2(x) - G\phi_1'(x)&&
\nonumber \\
{}+\frac{G\phi_2'(x)}{\omega + G\phi_2(x)}
\bigl[G\phi_1(x)+i(k+\beta'(x,\omega))\bigr]&=&0 
\hspace{1cm}
\label{betaeq}
\end{eqnarray}
subject to the boundary conditions $\beta(\infty,\omega)
=\beta^\prime(\infty,\omega)=0$.  It is this equation that we
solve numerically for a given background $\svec\phi(x)$ to determine
the phase shifts.

\subsection{Fermion One-Loop Contribution to the Effective Energy}

Having explained how to obtain the phase shifts for positive and negative
energies and parities, we next show how to compute the fermion quantum
corrections to the soliton energy following the method of
\cite{method1,method2,method3}.

We are interested in the ``effective energy'' of a classical field
configuration, which is the effective action per unit time for a time
independent field, and is the energy of the lowest state with a given
expectation value $\langle \svec\phi(x) \rangle$.  For a self-coupled
quantum field, as considered in Ref.~\cite{selfcoupled}, one introduces a
source that makes the configuration $\svec\phi(x)$ a stationary point of
the classical action.  A Legendre transformation then converts the energy
from a functional of the source to a functional of the classical background
field \cite{Cole}.  For the case at hand, however, we have suppressed the
quantum effects of the background field through the large-$N_F$ expansion. 
As a result, we can simply fix the classical backgound, at a cost in energy
$E_{\rm cl}$ that is given by the classical Hamiltonian, and then compute
the quantum corrections due to fermion fluctutatons in this background.

The fermion effective energy in the presence of the classical background
$\Delta E^{F}[\svec\phi\,]$ is given by the sum over fermion loop diagrams
with all sets of insertions of the background field.  These diagrams can be
evaluated by standard field theory techniques.  Since some of them will
diverge, we must introduce a regulator and counterterms that depend on the
regulator.  We will use dimensional regularization; other choices, such as
a momentum cutoff, should be equivalent (in general, we expect two
different regulators to lead to the same physical predictions as long as
they preserve the same symmetries).  The counterterms are then fixed by
renormalization conditions on the parameters of the theory.  Rather than
sum all the diagrams, our approach will be to write an equivalent
expression for the sum in terms of phase shifts, their Born approximations,
and a small number of renormalized diagrams (which in this case will happen
to be identically zero).  This transformation relies on the correspondence
between the Born expansion for the phase shifts and the diagramatic
expansion of the effective energy.  By comparing the two expressions for
the energy as analytic functions of the regulator, we will verify that in
making this correspondence, we have not introduced any finite errors or
changed the definition of the theory.

Formally, $\Delta E^{F}[\svec\phi\,]$ is also given by the shift in the
zero-point energies of the fermion modes, $-\fract{1}{2}\sum (|\omega| -
|\omega_0|)$, where $\{\omega_0\}$ are the eigenenergies in the free case. 
We rewrite this formal expression as a sum over bound states and an
integral over $k$ weighted by $\rho(k) -
\rho_0(k)$, the difference between the density of states in the free and
the interacting cases, giving
\begin{equation}
\Delta E^{F}[\svec\phi\,] =
-\Bigl(\fract{1}{2}\sum_j |\omega_j| + \fract{1}{2}\int_0^\infty
dk\,\sqrt{k^2 + m^2} (\rho(k)-\rho_0(k)) - \frac{m}{2} \Bigr)
+ \Delta E_{\rm ct} 
\label{efermion1}
\end{equation} 
where $\omega_l\in[-m,m]$ denote the discrete bound states.
The extra $m/2$ takes into account the contribution from the ``half-bound''
states at $\omega = \pm m$ in the free case
\cite{method1,method2,method3}.   $\Delta E_{\rm ct}$ is the contribution
from the counterterm Lagrangian, which is evaluated at tree level.  A
counterterm proportional to $\svec\phi\cdot\svec\phi$ is implicitly present in
eq.~(\ref{lagb}), which must be renormal\-ized due to the coupling to the
fermions.

Both the counterterms and the integral over the continuum are formally
infinite.  Since our model is renormalizable, the sum is finite.  In order
to avoid ambiguities associated with the separate divergent terms in
eq.~(\ref{efermion1}) we have extended the method of dimensional
regularization to this representation of the effective energy: 
Eq.~(\ref{efermion1}) and the subsequent development should be understood
as defined in $n$ dimensions, where all integrals are convergent and
all counterterms are finite.  All the manipulations that we perform in the
rest of this section are unambiguous in $n$ dimensions.  The final result
is then analytically continued back to 1+1 dimensions, where it
remains finite.  Since this extension of dimensional regularization is new
and may have wider application, we present it in a self-contained way in
Appendix~\ref{app:a}.

The difference between the free and interacting density of states
is given in terms of the scattering phase shifts by
\begin{equation}
\rho(k) - \rho_0(k) = \frac{1}{\pi}\frac{d}{dk} \delta_F(k)
\end{equation}
where
\begin{equation}
\delta_F(k)=\delta_+(\omega_{k})+\delta_+(-\omega_{k})
+\delta_-(\omega_{k})+\delta_-(-\omega_{k})
\label{totdel}
\end{equation}
and $\omega_{k}=+\sqrt{k^2 + m^2}$.  $\delta_{F}(k)$
sums the contributions from positive and negative energies and both
parities.  As in \cite{method1,method2,method3}, we use Levinson's theorem
\begin{eqnarray}
\delta_+(m) + \delta_+(-m) &=& \pi (n^+ - \fract{1}{2})\nonumber \\[1ex]
\delta_-(m) + \delta_-(-m) &=& \pi (n^- - \fract{1}{2})
\label{levin}
\end{eqnarray}
to relate the phase shifts at $k=0$ in each channel to the
number of bound states in that channel.  The factors of $\half$ are
peculiar to one dimension and reflect the existence of ``half-bound'' states at
threshold in the absence of interactions.  For a complete discussion, see
Refs.~\cite{method2} and \cite{Barton}.  We express Levinson's theorem as
\begin{equation}
0 = \half \sum_j m + \frac{m}{2\pi}\int_0^\infty dk
\frac{d}{dk} \delta_F(k) - \frac{m}{2}
\end{equation}
in order to rewrite
\begin{equation}
\Delta E^{F}[\svec\phi\,]=
-\Bigl(\fract{1}{2}\sum_j |\omega_j| + \frac{1}{2\pi}\int_0^\infty dk
\sqrt{k^2+m^2} \frac{d}{dk} \delta_{\rm F}(k) - \frac{m}{2} \Bigr)
+ \Delta E_{\rm ct}
\label{cas0}
\end{equation}
as
\begin{equation}
\Delta E^{F}[\svec\phi\,]=
-\Bigl(\fract{1}{2}\sum_j\left(|\omega_j|-m\right)
+\int_0^\infty \frac{dk}{2\pi}\left(\sqrt{k^2 + m^2} - m \right)
\frac{d}{dk}\delta_{\rm F}(k) \Bigl) +
\Delta E_{\rm ct}\,.
\label{cas1}
\end{equation}
This substitution, which is valid in arbitrary dimensions, will allow us to
formulate a Born expansion for eq.~(\ref{cas1}) without infrared divergences.

In terms of the shifted field $h = \phi_1 - v$, the $\svec\phi\cdot\svec\phi$
counterterm contribution to eq.~(\ref{cas1})~is
\begin{equation}
\Delta E_{\rm ct} = C \int (h^2 + 2 hv + \phi_2^2)\, dx  
\label{counter}
\end{equation}
consistent with the chiral symmetry of the fermion interaction.  We fix the
coefficient $C$ by the renormalization condition that the counterterm exactly
cancels the tadpole graph (the term linear in $h$), and we perform no
further finite renormalizations.  This choice fixes the counterterm
contribution to be equal to the tadpole graph plus those parts of the
graphs with two external lines that are related to the tadpole graph by
eq.~(\ref{counter}).  This contribution is entirely local, i.e., independent
of the external momenta.  In order to reexpress this contribution in terms
of phase shifts, we use the Born series.  We expand the phase shifts
in powers of the external fields $h$ and $\phi_2$.  The contribution from
the tadpole graph corresponds to the contribution from the first Born
approximation to the phase shift.  Expanding eq.~(\ref{beta}) and
eq.~(\ref{betaeq}) to lowest order in $h$ and $\phi_2$ we find that the
first Born approximation is given by
\begin{equation}
\delta^{(1)}_F(k) 
= \delta^{(1)}_+(\omega_{k})+\delta^{(1)}_+(-\omega_{k})
+ \delta^{(1)}_-(\omega_{k})+\delta^{(1)}_-(-\omega_{k})
= -\frac{4 G^2 v}{k} \int_0^\infty h(x)\, dx.
\end{equation} 
Once we have fixed the term linear in $h$, the
entire counterterm contribution to the phase shift is fixed by
eq.~(\ref{counter}) to be
\begin{equation}
\hat\delta_F(k)=\frac{2 G^2}{k}\int_0^\infty dx
\left(v^2-\svec\phi(x)^2\right) \, .
\label{born}
\end{equation}
Subtracting $\hat\delta_F(k)$ from $\delta_{\rm F}(k)$ in
eq.~(\ref{cas1}) then implements the full contribution of the counterterm,
giving
\begin{equation}
\Delta E^{F}[\svec\phi\,]=
-\fract{1}{2}\sum_j\left(|\omega_j| - m\right)
-\int_0^\infty \frac{dk}{2\pi}\left(\sqrt{k^2 + m^2} - m\right)
\frac{d}{dk}\left(\delta_{\rm F}(k)-\hat\delta_F(k)\right)\, .
\label{cas2}
\end{equation}

By expanding eq.~(\ref{beta}) and eq.~(\ref{betaeq}) for large $k$ we see
that $\hat\delta_F(k)$ gives the leading $1/k$ behavior of $\delta_{\rm
F}(k)$ for $k$ large, and thus the resulting integral over the continuum
is finite. We note that on the chiral circle $\svec\phi^2 = v^2$, the
counterterm contribution (\ref{born}) to the energy vanishes, implying that
the one-loop quantum contribution to the energy is finite.  This is a
consequence of the chiral invariance of the boson-fermion interaction,
which forces the counterterms involving the ``would-be'' Goldstone boson
to contain at least two derivative operators.  On dimensional grounds these
counterterms must be finite in 1+1 dimensions.  In the present context,
it implies that $\delta_{\rm F}(k)$ decreases more rapidly than $1/k$ as
$k\to\infty$ for scalar configurations on the chiral circle, and our
numerical computation indicates that in this case $\delta_{\rm F}(k)$ goes
like $1/k^3$ for $k$ large.

For numerical computations, it is convenient to integrate eq.~(\ref{cas2})
by parts, yielding
\begin{equation}
\Delta E^{F}[\svec\phi\,] = -\fract{1}{2}\sum_j\left(|\omega_j|-m\right)
+\int_0^\infty \frac{dk}{2\pi}\frac{k}{\sqrt{k^2+m^2}}
\bigl(\delta_{\rm F}(k)-\hat\delta_F(k)\bigr)\, .
\label{cas3}
\end{equation}
This is the finite expression for the fermion one-loop contribution to the
effective energy which we evaluate numerically.  The analysis of the first part
of this section gives us the phase shifts, and we find the bound states by
ordinary shooting methods, with Levinson's theorem telling us how many to
look for.

\subsection{The Fermion Number of a Configuration}

We are interested in fermionic solitons and therefore must keep track of
the fermion number. If, as we adiabatically turn on the background
configuration starting from the trivial vacuum, a fermion level crosses
zero from above, the vacuum will have this state filled, giving the
configuration a positive fermion number. (If the crossing is in the other
direction, there is a negative fermion number.)  The fermion energy levels
are $N_{F}$-fold degenerate, so the configuration carries each of the
$N_{F}$ fermion numbers. We know from general considerations that if
$\svec\phi(x)$ circles $\svec\phi=(0,0)$ as $\svec\phi$ goes from
$(v,0)$ at  $x=-\infty$ to $(v,0)$ at $x=\infty$, then the vacuum state
will carry nonzero fermion number provided that $w$, the scale over which
$\svec\phi$ varies, is much larger than the fermion Compton wavelength $1/m$
\cite{Farhi,GoldWil}.  

In Ref.~\cite{tbaglevi} we show how to compute this fermion number directly
using Levinson's theorem, which relates the phase shifts at $\omega = \pm
m$ to the total number of bound states.  We use that the difference between
the phase shift at the positive energy threshold, $\omega=m$, and the actual
number of positive (negative) energy bound states $n_>~(n_<)$ counts the
fermion number ${\cal Q}$ of the vacuum:

\begin{eqnarray}
{\cal Q}[\svec\phi\,] &=&
N_F \Bigl(\frac{1}{\pi}\bigl(\delta_{+}(m)+\delta_{-}(m)\bigr) 
+\fract{1}{2} - n_{>} \Bigr) \nonumber\\
&=&- N_F \Bigl(\frac{1}{\pi}\bigl(\delta_{+}(-m)+\delta_{-}(-m)\bigr)
+\fract{1}{2} - n_ {<} \Bigr) \, .
\label{qvac}
\end{eqnarray}
As noted above, the extra $\half$ is peculiar to one dimension, and is
essential to get correct (integer) values for ${\cal Q}$.  ``Half-bound''
states at the respective thresholds count as $\fract{1}{2}$ in $n_>$ or
$n_<$. The configurations that we consider circle $\svec\phi=(0,0)$ at
most  once, so ${\cal Q}$ is either $0$ or $N_F$. 

We are interested in states with fermion number $N_F$.  If ${\cal Q}=N_F$,
then the state we want is the vacuum.  If ${\cal Q}=0$, then we build the
lowest energy state with fermion number $N_F$ by filling the lowest
positive energy level $\omega_1$ of eq.~(\ref{dirham}) with $N_F$
fermions.  Thus the total fermionic contribution to the energy is
\begin{equation}
E^F = \Delta E^F + \omega_1(N_F - {\cal Q}) \, .
\end{equation}

We note that $E^F$ varies {\it smoothly} with $\omega_{1}$ even as
$\omega_{1}$ crosses zero.  The vacuum and valence contributions each have
discontinuities in slope when $\omega_{1}=0$, but the contribution from
this level is $-\half|\omega_{1}|$ from the vacuum energy in
eq.~(\ref{cas3}) plus $\omega_{1}\theta(\omega_{1})$ as a filled valence
level.  The sum is $\half\omega_{1}$ for all $\omega_{1}$, which is smooth.

The spectrum also contains states with total charge between zero and
$N_F$.  These states are constructed by an appropriate filling of the
tightly bound level if $\omega_1>0$ or the emptying of that level if
$\omega_1<0$.  We do not consider such states because their binding will be
smaller than that of states with charge $N_F$.  We note, however, that if
the effects we find in the large $N_F$ limit persist for moderate values of
$N_F$, then in that case a stable soliton with fermion number $1$ would
represent an elementary fermion.

It will also be helpful to consider the fermion charge density. The vacuum
contribution is given by summing over the eigenstates obtained from
eq.~(\ref{sdef}),
\begin{equation}
\langle\Omega|\Psi^{\dagger}(x)\Psi(x) |\Omega\rangle =
-N_F \int_{-\infty}^{\infty} {\rm sgn}(\omega) \half  \Bigl(
\psi_+^\omega(x)^\dagger \psi_+^\omega(x) +
\psi_-^\omega(x)^\dagger \psi_-^\omega(x) \Bigr)\, d\omega
\label{qdensity0}
\end{equation}
where the integral over $\omega$ includes both continuum and bound
states. We have restored the suppressed $\omega$ label on the
wavefunctions and normalized them by
\begin{equation}
\int \psi_\pm^\omega(x)^\dagger \psi_\pm^{\omega'}(x) \, dx = \delta(\omega
- \omega')
\end{equation}
where the right hand side is interpreted as a Dirac delta function for
continuum states and a Kronecker delta for bound states.  The norm of the
scattering states is related to the phase shifts by
\begin{equation}
\int \left(\psi_\pm^\omega(x)^\dagger \psi_\pm^\omega(x) - 1\right) dx
= \frac{1}{\pi} \frac{d \delta_\pm(\omega)}{d\omega}
\qquad ({\rm for}\, |\omega| > m)
\end{equation}
as derived in \cite{method2}.  The spatial integral over the bound
state contribution to (\ref{qdensity0}) yields the difference between the
numbers of positive and negative energy bound states, $n_>-n_<$.  Thus by
integrating eq.~(\ref{qdensity0}) over $x$ and using Levinson's theorem and
the fact that the total number of states is unchanged, we obtain
eq.~(\ref{qvac}).

If ${\cal Q}=0$ we have to explicitly add the contribution of the level
$\psi_1(x)$ with the lowest positive eigenvalue.  Thus the total charge
density is given by
\begin{equation}
j_0(x)=\left(N_F- {\cal Q}\right)\psi_1^{\dagger}(x)\psi_1(x)
+\langle\Omega| \Psi^{\dagger}(x)\Psi(x) |\Omega\rangle \,.
\label{qdensity}
\end{equation}

\section{The Total Energy}\label{sec:4}

Now that the theory has been prescribed by
${\cal L}_{B}+{\cal L}_{F}$, it is useful to introduce dimensionless
variables to simplify our analysis.  We measure all energies in units
of the perturbative fermion mass, $m=G v$, and all
distances in terms of $\xi=mx$.
We are interested in calculating the energy of a fixed background
configuration $\svec\phi(x)$, which we describe in terms
of a small set of dimensionless variational parameters $\{\zeta_i\}$.
Next we rescale $\svec\phi$ by $v$ (both of which are
dimensionless in one dimension) so that $\svec\phi\to(1,0)$ as
$|\xi|\to\pm\infty$, and we define dimensionless couplings
\begin{equation}
\tilde{\alpha}=\frac{\alpha}{G^2}
\qquad {\rm and} \qquad
\tilde{\lambda}=\frac{\lambda}{G^2} \, .
\label{workvar}
\end{equation}
With all these rescalings the boson classical energy can
be written as
\begin{eqnarray}
\frac{E_{\rm cl}[\svec\phi\,]}{m}&=&
 v^2\int_{-\infty}^\infty d\xi\,
\biggl\{ \fract{1}{2} \svec\phi\,'\cdot\svec\phi\,'+
\frac{\tilde{\lambda}}{8} \Bigl(\svec\phi\cdot\svec\phi
-1+\frac{2\tilde{\alpha}}{\tilde{\lambda}}\Bigr)^2
-\frac{\tilde{\lambda}}{2}
\Bigl(\frac{\tilde{\alpha}}{\tilde{\lambda}}\Bigr)^2
-\tilde{\alpha}(\phi_1-1)\biggr\}
\nonumber \\[1ex]
&=& v^2{\cal E}_{\rm cl}(\tilde{\alpha},\tilde{\lambda},\{\zeta_{i}\})
\label{ecl}
\end{eqnarray}
where prime denotes a derivative with respect to $\xi$.

The fermion one-loop contribution to the energy is given by a regulated sum
over the absolute values of the eigenfrequencies, $\{\omega_{n}\}$, of
$H[\svec\phi\,]$.  Once we measure distances in units of $\xi$ and choose
dimensionless variational parameters $\{\zeta_{i}\}$, the perturbative fermion
mass $m$ factors out of the Dirac Hamiltonian, so its eigenvalues
$\{\omega_{n}\}$ scale with $m$.  In all, the fermion one-loop vacuum energy
can be written as
\begin{equation}
E^{F}[\svec\phi\,]=N_{F}m{\cal E}^{F}(\{\zeta_{i}\})\, .
\label{f1loop}
\end{equation}

Thus we compute
\begin{equation}
\frac{E_{\rm tot}[\svec\phi\,]}{N_F\, m}=
\frac{ v^2}{N_F}
{\cal E}_{\rm cl}(\tilde{\alpha},\tilde{\lambda},\{\zeta_i\})
+{\cal E}^{F}(\{\zeta_i\}) \, .
\label{loopcount}
\end{equation}
This computation is exact in the limit $N_F\to\infty$ and
$v^2\to\infty$ with the ratio $v^2/N_F$ held fixed, since the contributions
we have neglected all arise from loops with internal bosons, which are
suppressed by $v^2$ relative to the classical energy and by $N_F$ relative
to the one-loop fermion energy.

To compare the energy of our background configuration to the energy of the
state with the same charge built on top of the translationally invariant vacuum,
we study the effective energy per fermion minus $N_F$ times the perturbative
fermion mass $G v$, which we denote as  ${\cal B}[\svec\phi\,]$.  In terms of
rescaled variables,
\begin{equation}
{\cal B}\Bigl(\frac{v^2}{N_F},\tilde{\alpha},\tilde{\lambda},\{\zeta_i\}\Bigr) =
\frac{ v^2}{N_F} {\cal E}_{\rm cl}\bigl(\tilde{\alpha}, \tilde{\lambda},
\{\zeta_i\}\bigr) + {\cal E}^{F}(\{\zeta_i\})- 1 \, .
\label{etot}
\end{equation}
A field configuration with ${\cal B}<0$ is energetically favored over
the state with the same charge $N_F$ built on top of the translationally
invariant vacuum.

\section{Numerical Explorations}\label{sec:5}

In this section we present the results of our numerical studies. We will
show that the model has a stable fermionic soliton for a wide range of the
model parameters $\tilde\lambda$, $\tilde\alpha$, and $v^2/N_{F}$. For
fixed model parameters, we search over the variational parameters looking
for a bound solitonic fermion.  The classical contribution to the soliton
energy, ${\cal E}_{\rm cl}$ in eq.~(\ref{etot}), is positive and is scaled
by $v^2/N_{F}$.  The fermion one-loop contribution, ${\cal E}^{F}-1$, is
generically negative.  Thus, the existence of a stable fermionic soliton
can be discussed in terms of a maximum value of $v^{2}/N_{F}$ for a given
choice of $\tilde\lambda$ and  $\tilde\alpha$.

\subsection{Variational Ansatz}

We choose an \emph{ansatz} $\svec\phi_{I}(\xi)$ characterized by the 
parameters $R$ and $w$,
\begin{equation}
\svec\phi_{I}(\xi,R,w) = \bigl(1-R+R \cos\Theta_{I}(\xi,w), R
\sin\Theta_{I}(\xi,w)\bigr)
\label{ansatz1}
\end{equation}
with
\begin{equation}
\Theta_{I}(\xi,w) = \pi\bigl(1+\tanh(\xi/w)\bigr).
\label{ansatz2}
\end{equation} 
For fixed $R$ and $w$, $\svec\phi_{I}$ describes a circle centered at
$1-R$ with radius $R$ as $\xi$ varies.  $w$ gives the characteristic size
of the configuration.  Varying $R$ allows us to interpolate between the
trivial vacuum, $\svec\phi = (1,0)$, and a configuration that circles the
origin once while staying on the chiral circle, $|\svec\phi| = 1$.  
Configurations with $R>1$ go beyond the chiral circle.  When the symmetry
breaking parameter $\tilde\alpha$ is large,  $\svec\phi=(-1,0)$ is
energetically disfavored, leading to a preference for $R$ to go to zero to
avoid that point.

We have also considered other \emph{ans\"atze}, but we found that they
yielded variational minima that had comparable or higher energy.  Thus we
will restrict our attention to this choice.

\subsection{Stability Studies}

To search for the soliton, we must compute the classical energy
${\cal E}_{\rm cl} (\tilde\alpha,\tilde\lambda,R,w)$ and the fermionic
contribution ${\cal E}^{F}(R,w)$.  Our scaling has simplified the
numerical \emph{ansatz} by putting all the dependence on the model parameters
$\tilde\lambda$, $\tilde\alpha$, and $ v^{2}/N_{F}$ into the
classical energy, which is easy to compute.  The fermionic
contribution, which is harder to compute, depends only on the
variational parameters $R$ and $w$. 

\begin{figure}[hbt]
\centerline{
\BoxedEPSF{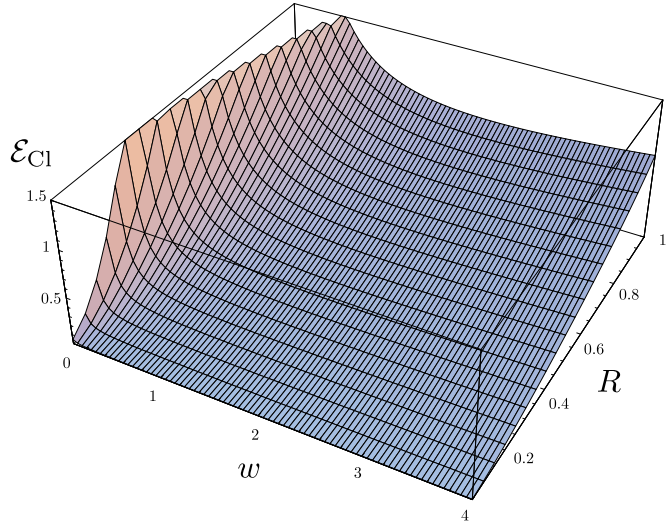 scaled 500}}\bigskip
\caption{\sl The classical energy,
${\cal E}_{\rm cl}$, as a function of the variational parameters $R$ and $w$
in eq.~(\protect\ref{ecl}).  The parameters of the bosonic Lagrangian are
$\tilde{\lambda}=1.0$ and $\tilde{\alpha}=0.5$. }
\label{classical}
\end{figure}

The classical energy is computed from eq.~(\ref{ecl}). A typical energy 
surface for \emph{ansatz} $\phi_{I}$ is shown in Fig.~\ref{classical} as a 
function of $R$ and $w$ for a generic choice of model parameters, 
$\tilde\lambda$ and $\tilde\alpha$.
Simple scaling arguments allow us to understand the principal
features of Fig.~\ref{classical}.  For fixed $w$, ${\cal E}_{\rm
cl}(\tilde\alpha,\tilde\lambda,R,w)$ vanishes as $R\to 0$ because the
\emph{ansatz} approaches the classical vacuum for all $\xi$.  For fixed $R\ne 0$,
${\cal E}_{\rm cl}(\tilde\alpha,\tilde\lambda,R,w)$ diverges like $1/w$ for
small $w$ due to the $|\svec\phi'|^{2}$ terms in eq.~(\ref{ecl}).  For
large $w$, ${\cal E}_{\rm cl}(\tilde\alpha,\tilde\lambda,R,w)$ grows
linearly with $w$ for fixed $R\ne1$ because  $V(\svec\phi)$ is nonzero in a
region of size $w$.  

\begin{figure}[hbt]
\centerline{
\BoxedEPSF{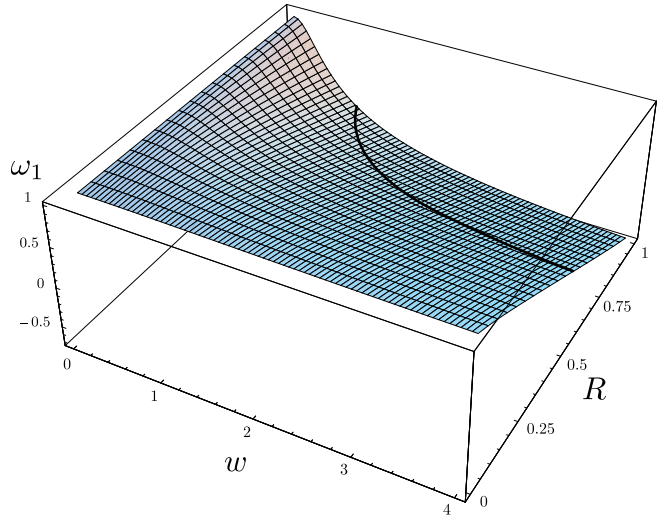 scaled 500}}\bigskip
\caption{\sl The lowest quark eigenenergy, $\omega_{1}$, for \emph{ansatz}
$\phi_{I}$, as a function of $R$ and $w$.  Note that for large $R$ and
$w$, $\omega_{1}$ is negative. A solid curve marks the contour $\omega_1 =
0$.}
\label{valence}
\end{figure}

Next we turn to the fermion contribution to the energy.  There are two
distinct contributions discussed in the previous section: first, the energy
of any filled ``valence'' level, and second the vacuum energy.  An
important property of our \emph{ansatz} is its propensity to bind a fermion.  This
can be seen in Fig.~\ref{valence}, where we plot the lowest fermion
eigenenergy $\omega_{1}$ as a function of the variational parameters $R$
and $w$ for the \emph{ansatz} $\phi_{I}$.  Since we are using scaled variables the
fermion continuum begins at $\omega=1$.  Notice that for $R$ near 1, which
puts $\phi_{I}$ close to the chiral circle, the lowest fermion
eigenenergy decreases quickly with increasing $w$.  Even a modest value of
$w$ leads to a {\it negative} eigenenergy.

The strongly bound fermion state drives the formation of a fermionic
soliton.  The situation is reminiscent of the appearance of a fermion zero
mode in the presence a soliton in a theory of a real scalar
field\cite{JR}.  It is expected from considerations of level crossings in
the presence of an adiabatically changing field: If we fix $R>\half$ so
that the orbit in the $(\phi_{1},\phi_{2})$ plane circles the origin, then
as $w$ increases from $0$ to $\infty$ one fermion level must cross from the
positive to the negative continuum \cite{Farhi,GoldWil}.  Our numerical
studies indicate that $\omega_{1}$ is never negative unless the orbit of
$\svec\phi$ circles the origin.

\begin{figure}[hbt]
\centerline{
\BoxedEPSF{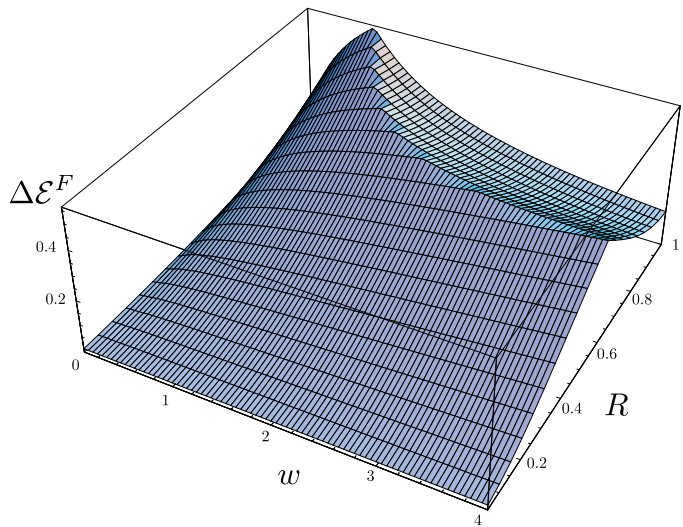 scaled 500}}\bigskip
\caption{\sl The vacuum contribution to the 
one-loop fermion energy as a function of $R$ and $w$.  The sharp edge 
in this contribution occurs when the lowest fermion eigenvalue 
crosses zero.  See Fig.~\ref{valence}.}
\label{casimir}
\end{figure}

In Appendix~\ref{app:b}, we explore the generality of level crossings for field
configurations on the chiral circle.  We show that the WKB approximation
yields a single strongly bound level in the positive parity channel, which
crosses from the positive to negative continuum as the width is
increased.  Note that merely examining the spectrum in the limit of large
width gives no evidence of the level crossing.  It is essential to follow
the levels as a function of width or use the Levinson's theorem method,
eq.~(\ref{qvac}).

\begin{figure}[hbt]
\centerline{
\BoxedEPSF{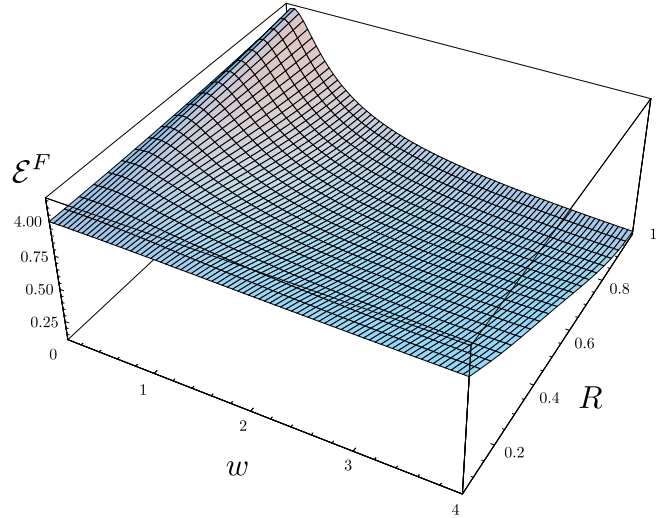 scaled 500}}\bigskip
\label{fig_4}
\caption{\sl The fermion one-loop contribution, ${\cal E}^{\rm F}(R,w)$, 
to the energy of a state of fermion number $N_F$ for the \emph{ansatz} $\phi_{I}$.}
\label{fermitot}
\end{figure}

Next we calculate the contribution from the fermionic vacuum -- the
Casimir energy -- given by eq.~(\ref{cas3}).  Note that this sum over the
eigenenergies of the Dirac Hamiltonian is of the same order in $\nff$ as the
energy of the lowest eigenmode and must be included in a consistent
calculation of the one-loop fermion contribution to the effective energy. 
This piece of the energy is also a function only of the \emph{ansatz} parameters
$R$ and $w$.  It is shown in Fig.~\ref{casimir}. It displays a
discontinuity in slope at the point where the lowest eigenvalue passes
through zero.  Fig.~\ref{fermitot} shows that the sum of the Casimir
energy and the contribution from $\omega_1$ is smooth, in agreement with
the discussion in the previous section.

\begin{figure}[hbt]
\centerline{
\BoxedEPSF{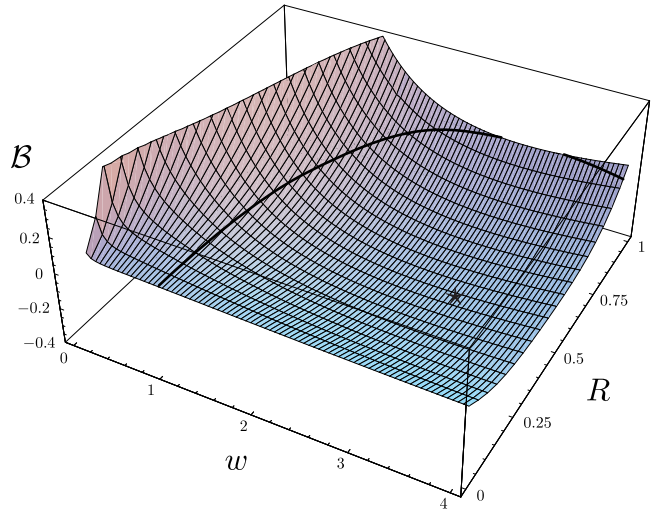 scaled 500}}\bigskip
\caption{\sl ${\cal B}$  as a function of the \emph{ansatz} parameters for $\tilde
\alpha = 0.5$, $\tilde \lambda = 1.0$, and $v/\sqrt{N_F} = 0.375$. A solid
curve marks the contour
${\cal B} = 0$, and a star indicates the minimum at
$w=2.808$ and $R=0.586$.}
\label{example}
\end{figure}

\begin{figure}[hbt]
\centerline{
\BoxedEPSF{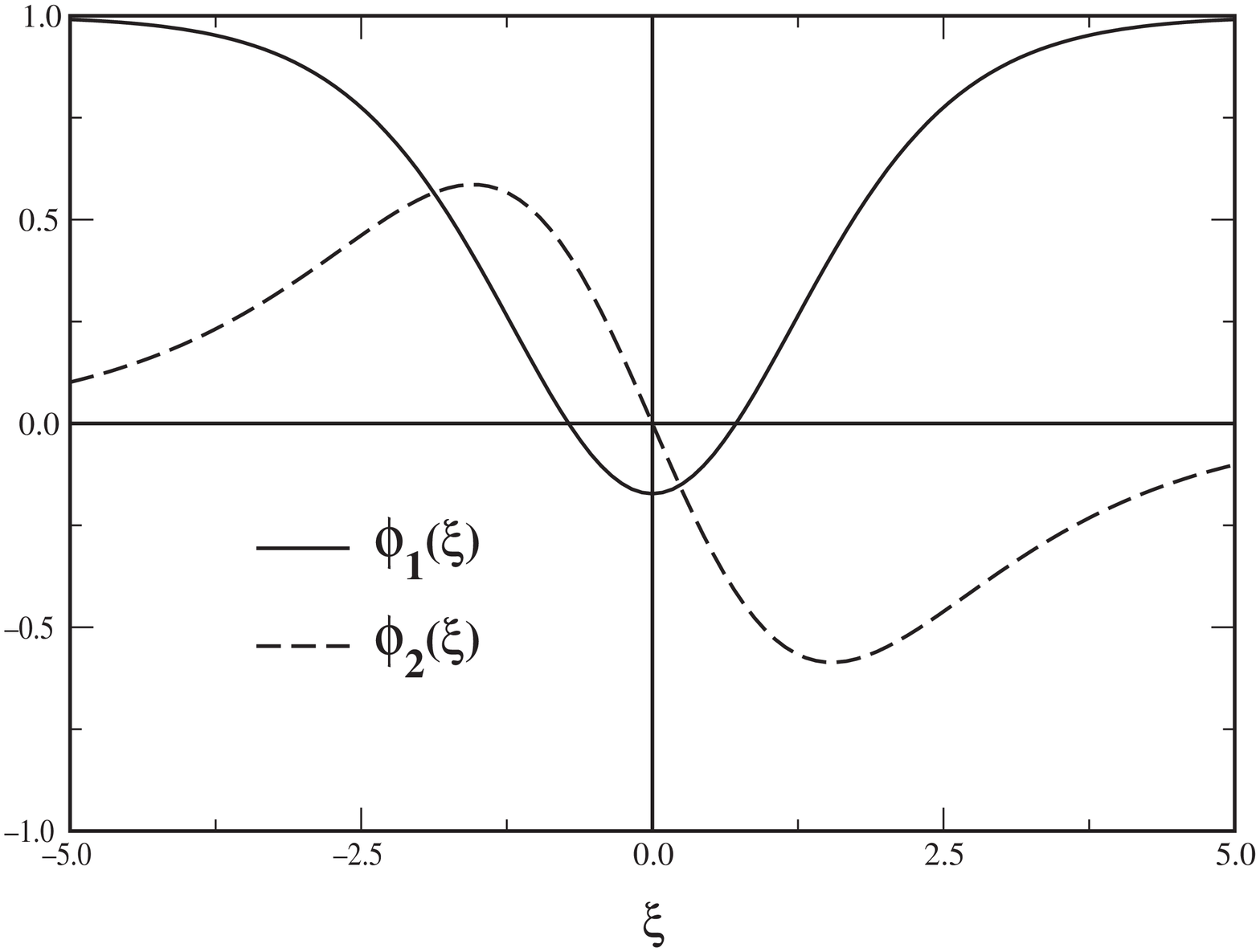 scaled 300}
\quad
\BoxedEPSF{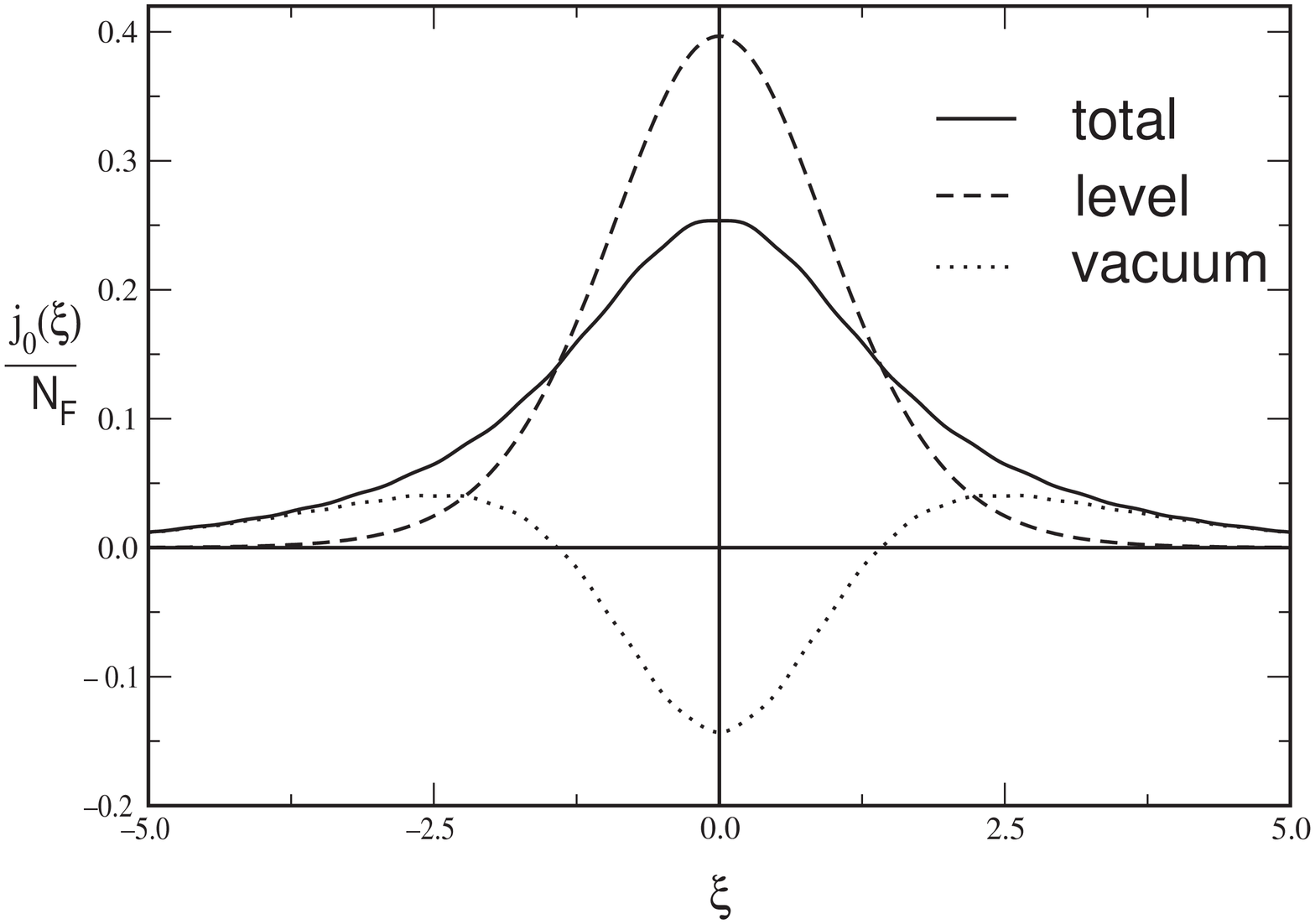 scaled 300}}
\vskip0.5cm
\caption{\sl $\phi_{1}$, $\phi_2$, and the fermion number density $j_0$ at
the variational minimum for
$\tilde\alpha = 0.5$,  $\tilde\lambda = 1.0$, and $v/\sqrt{N_F} = 0.375$,
which is at 
$R = 0.586$,
$w = 2.808$.  The left panel shows $\phi_1(\xi)$ and $\phi_2(\xi)$ at
this point, and the right panel shows $j_0(\xi)$, which gets contributions from
both the fermion vacuum and the filled valence level.}
\label{soliton}
\end{figure}

Finally, we combine the classical energy and the one-loop fermion energy to
form ${\cal B}(v^{2}/N_F,\tilde{\alpha},\tilde{\lambda},\{\zeta_i\})$ and
search for a stable fermionic soliton.  As an example of our results,
Fig.~\ref{example} shows ${\cal B}$ as a function of $R$ and $w$ for
$\tilde\lambda=1$, $\tilde\alpha=0.50$, and $ v/\sqrt{N_{F}}=0.375$. ${\cal
B} < 0$ is the signal of a stable fermionic soliton.  The domain of
negative ${\cal B}$ is large and the variational approximation to the
fermionic soliton lies at the minimum: $R=0.586$, $w = 2.808$.  In this
case the lowest fermion eigenenergy, $\omega_{1}=0.0985$, is slightly 
positive so this level must be filled. The variational approximation to 
the binding energy is $0.253$. In Fig.~\ref{soliton} we show $\phi_{1}(\xi)$, 
$\phi_{2}(\xi)$, and $j_0(x)$. We see that the charge density is
well localized around the center of the soliton.

\begin{figure}[hbt]
\centerline{
\BoxedEPSF{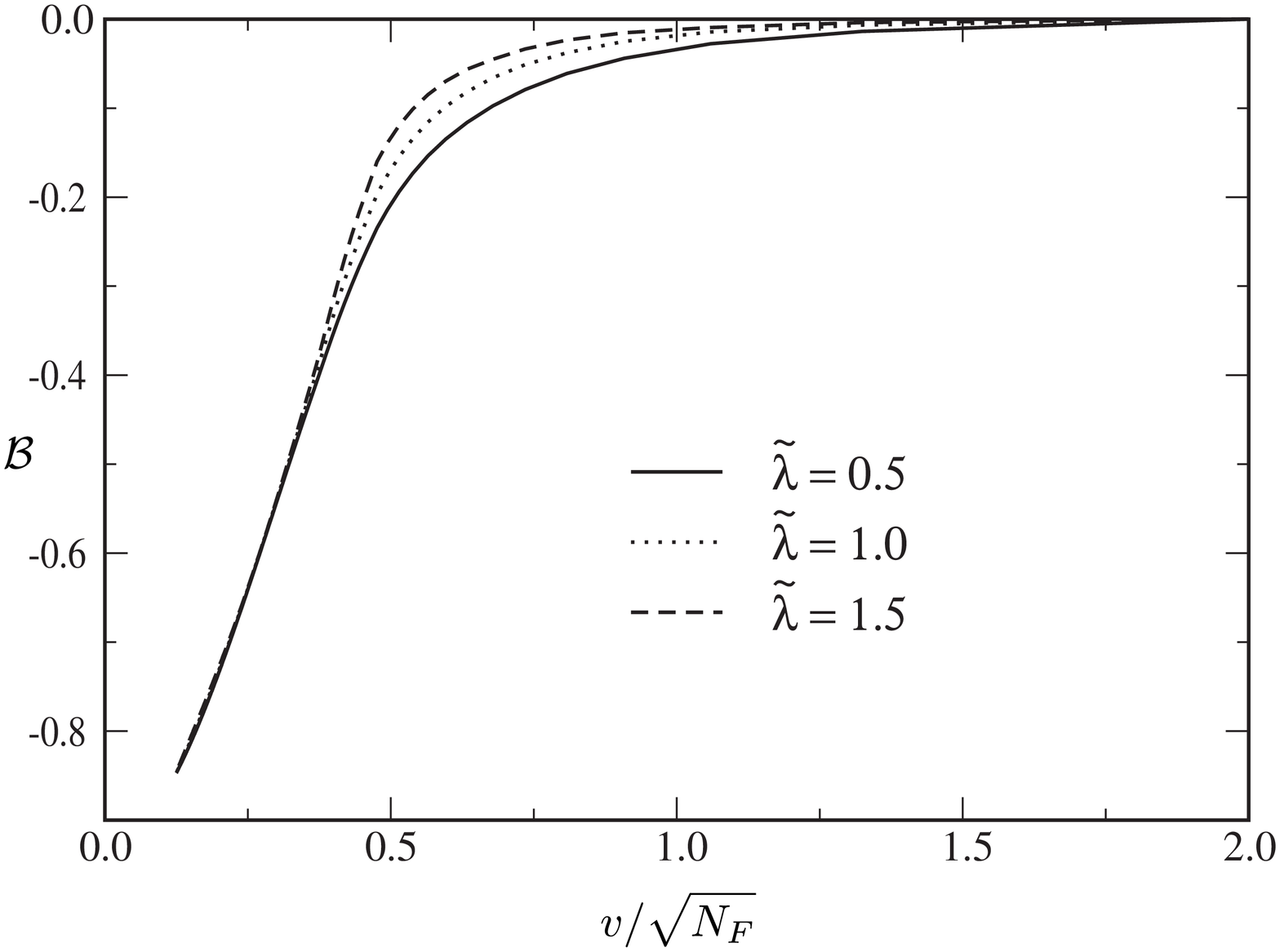 scaled 300}
\quad
\BoxedEPSF{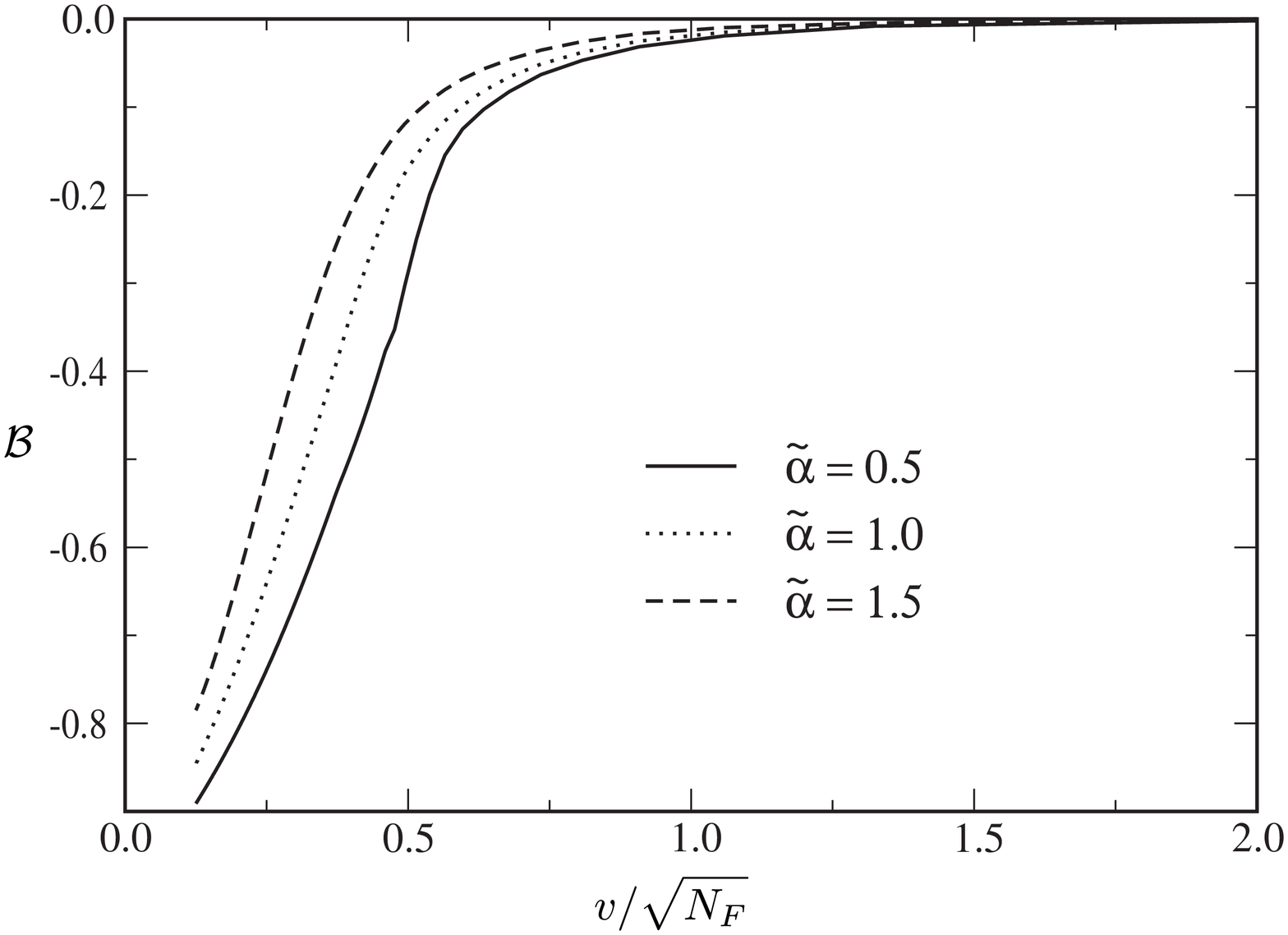 scaled 300}}
\vskip0.5cm
\caption{\sl ${\cal B}$ as a function of $v/\sqrt{N_F}$ for
various values $\tilde{\lambda}$ with $\tilde{\alpha}=0.25$ (left panel)
and for various values $\tilde{\alpha}$ with $\tilde{\lambda}=1.0$ (right
panel).}
\label{alphalambda}
\end{figure}

We are now in a position to study the stability of the fermionic soliton
as a function of the model parameters $\tilde\lambda$, $\tilde\alpha$, and
$ v^{2}/N_{F}$.  For each choice of parameters we minimize ${\cal B}$ over
the \emph{ansatz} parameters, $R$ and $w$.  The output is ${\cal
B}(\tilde\lambda,\tilde\alpha, v^{2}/N_{F})$ and the parameters
$R(\tilde\lambda,\tilde\alpha, v^{2}/N_{F})$ and
$w(\tilde\lambda,\tilde\alpha, v^{2}/N_{F})$ that minimize ${\cal B}$. 
In Fig.~\ref{alphalambda} we plot ${\cal B}$ versus 
$v/\sqrt{N_{F}}$ for various choices of $\tilde\lambda$ and
$\tilde\alpha$ and see that binding occurs over a wide range of model
parameters.  Binding is a quantum phenomenon -- it is maximal when $
v/\sqrt{N_{F}}$, which multiplies the classical contribution, vanishes, and
decreases with increasing $ v/\sqrt{N_{F}}$.

\begin{figure}[hbt]
\centerline{
\BoxedEPSF{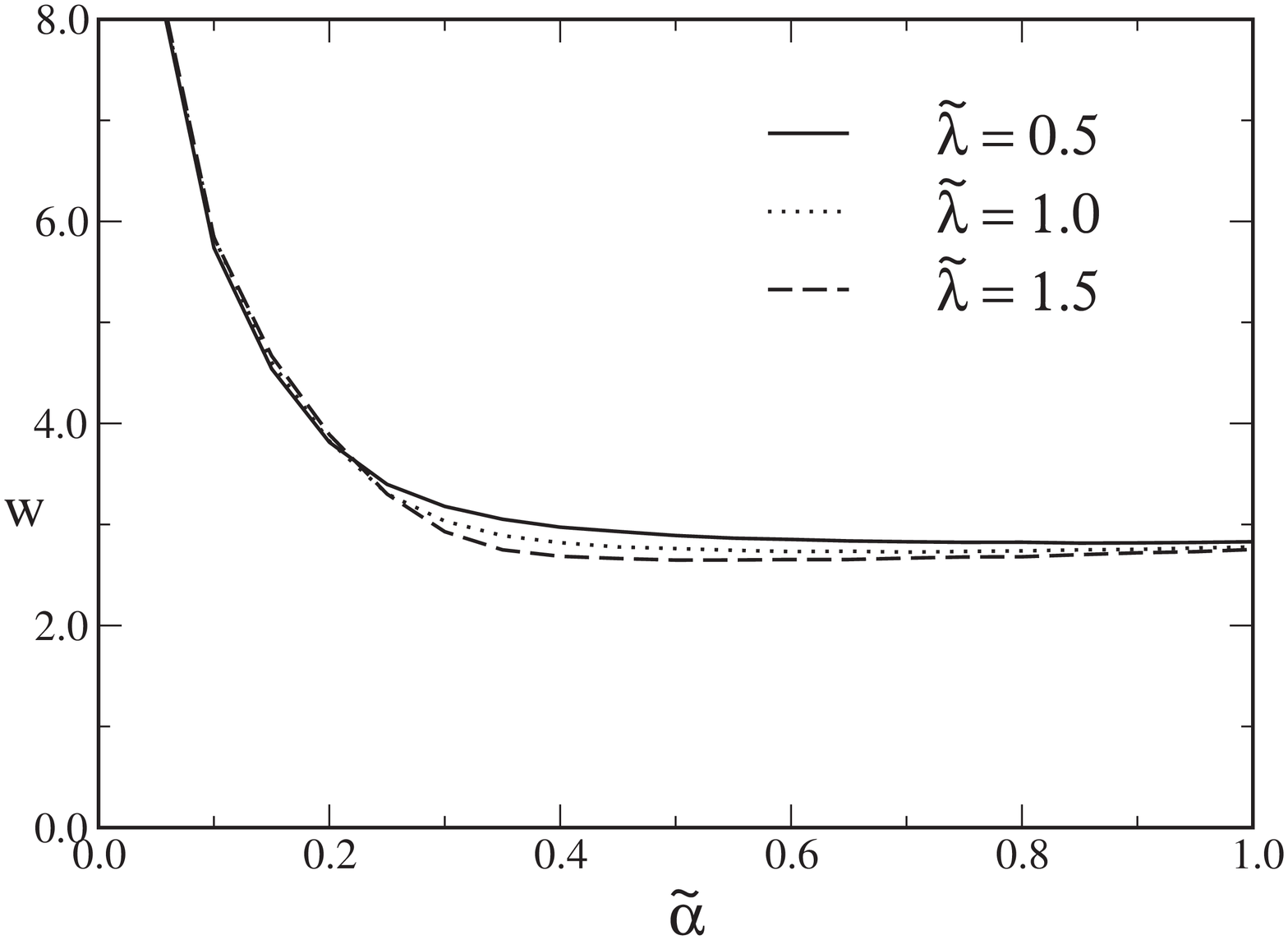 scaled 300}
\quad
\BoxedEPSF{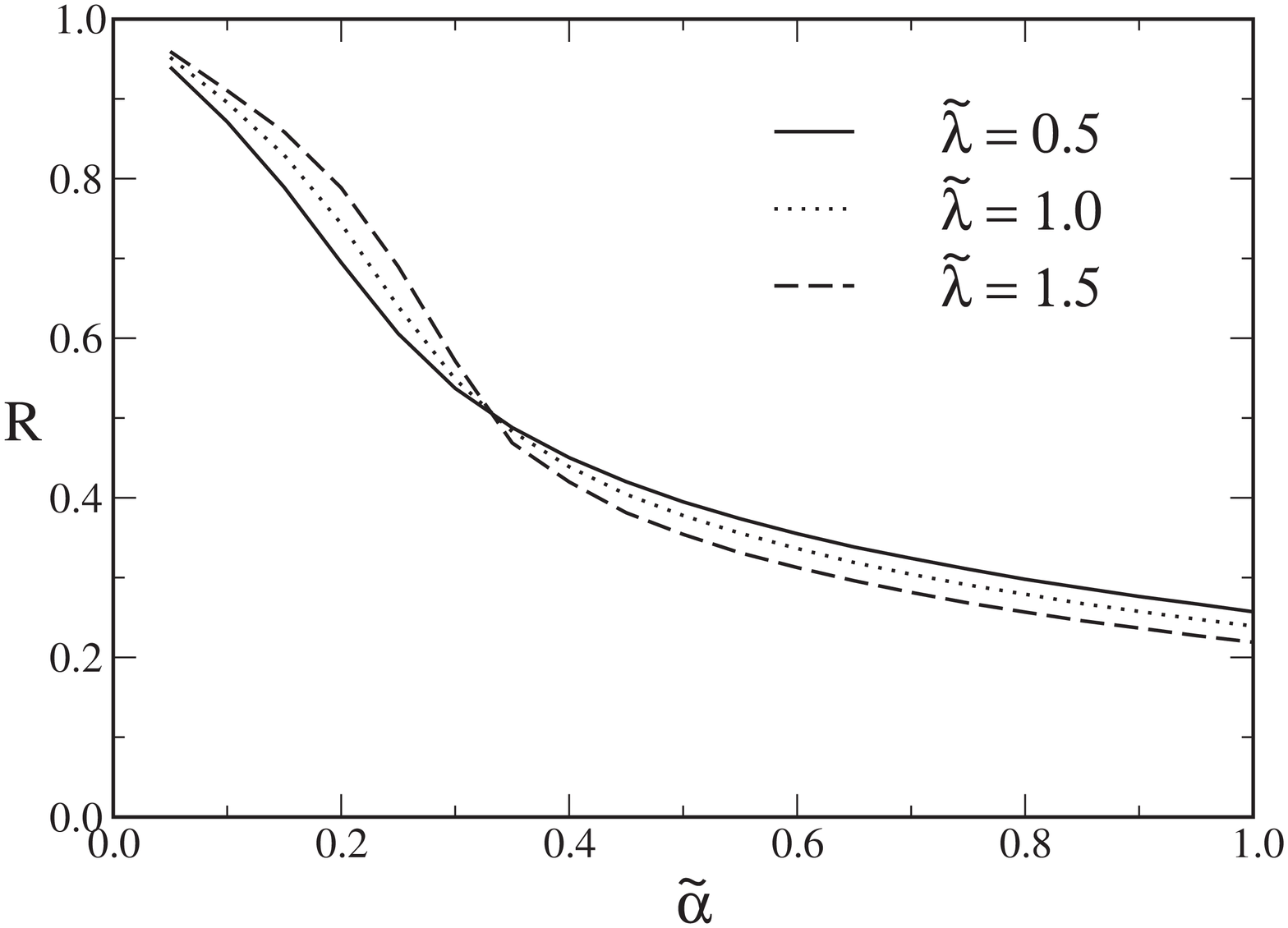 scaled 300}}
\bigskip
\caption{\sl The width   (\emph{left panel})
and the radius   (\emph{right panel}) of the
configurations that minimize the total energy
as a function of the explicit symmetry breaking $\tilde{\alpha}$.
Several values of the Higgs coupling constant $\tilde{\lambda}$
are considered and $v/\sqrt{N_F}=0.375$.}
\label{widthradius}
\end{figure}

\begin{figure}[hbt]
\centerline{\BoxedEPSF{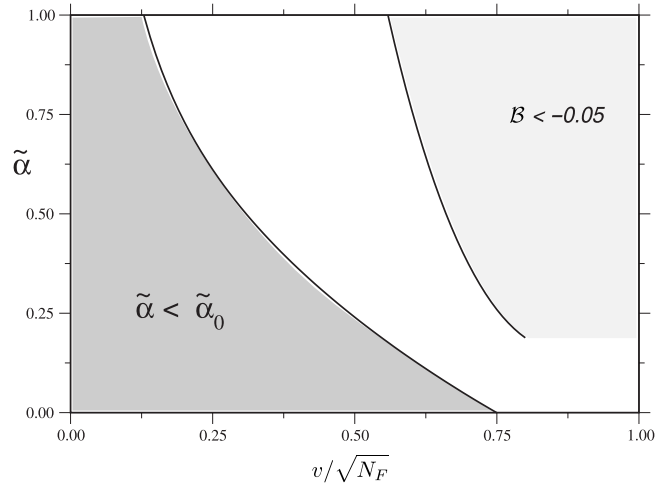 scaled 350}}
\bigskip
\caption{\sl The regions of soliton stability in the plane of $v/\sqrt{N_F}$ and
$\tilde\alpha$.  In the shaded area on the left, a growing width
indicates potential infrared instabilities.  In the shaded area on the right, the
soliton is bound by less than 5 percent.  In between, we have a stable, tightly
bound soliton.}
\label{critical}
\end{figure}

The dependence on the explicit symmetry breaking parameter $\tilde\alpha$
is particularly important because of the infrared problems we expect to arise
as $\tilde\alpha\to 0$.  Fig.~\ref{alphalambda} confirms our expectation that
${\cal B}\to -1$ (strong binding) as $\tilde\alpha\to 0$.  However this is a
suspect limit.  A look at the width of the soliton displays the problem.  This is
shown in Fig.~\ref{widthradius} where $w$ and
$R$ are plotted versus $\tilde\alpha$ for various choices of
$\tilde\lambda$ and a typical value of $v/\sqrt{N_{F}}$. Two qualitatively
different regimes are separated at $\tilde\alpha_{0}\approx 0.3$.  Above
$\tilde\alpha_{0}$ the size of the soliton is almost independent of
$\tilde\alpha$.  Below $\tilde\alpha_{0}$ the width of the soliton diverges like
$1/\sqrt{\tilde\alpha}$ and the \emph{ansatz} moves towards the chiral
circle, $R=1$.  However, this is where we expect the model to be invalidated by
infrared divergences generated by the would-be Goldstone mode.  Therefore
we only trust our results in the domain
$\tilde\alpha>\tilde\alpha_{0}$.  We have made plots analogous to
Fig.~\ref{soliton} for various choices of $\tilde\lambda$ and $v/\sqrt{N_{F}}$. 
We find that $\tilde\alpha_{0}$ depends weakly on
$\tilde\lambda$ but strongly on $\nf$.  In Fig.~\ref{critical} we plot
$\tilde\alpha_{0}(\nf)$ defined as the point where the break occurs in figures
analogous to Fig.~\ref{soliton}.  In the absence of a complete analysis of the
infrared problem as $\tilde\alpha\to 0$, we only trust our results when
$\tilde\alpha>\tilde\alpha_{0}(\nf)$.

For fixed $\alpha$, when $\nf$ is too small we run into the infrared
instabilities we described above.  When $\nf$ is very large the soliton
binding becomes small.  The interesting domain is between these two
extremes where a stable soliton can be reliably described.  To
indicate this domain we have added a contour corresponding to 5\%
binding in Fig.~\ref{critical}.  In between we have found a
strongly bound fermionic soliton.

\section{Discussion and Conclusions}\label{sec:6}

We have shown that quantum effects can stabilize a soliton in a theory with
no soliton at the classical level.  In order to convincingly demonstrate
this phenomenon, it is necessary to consider a renormalizable model.  We
study a (1+1)-dimensional model with a chirally invariant Yukawa
interaction between fermions and scalars.  The scalar self-interaction is
given by a Higgs potential with an explicit symmetry breaking term.
(Without the explicit symmetry breaking, the fluctuations of the Goldstone
modes would restore the spontaneously broken symmetry.)  We define the
renormalized energy functional using counterterms determined in the
perturbative sector of the theory.  Our phase shift formalism allows us to
unambiguously include the exact fermion one-loop quantum corrections in a
manner that is computationally tractable.  The omission of the scalar loop
contribution is justified by assuming a large number of fermion flavors.

Holding the total fermion number fixed, we performed a variational
calculation to search for configurations with lower energy than the same
number of free fermions.  Once we have found a bound configuration we are
assured that a stable soliton exists since its energy can only be smaller
than the variational minimum.  As expected, when the typical extension of
the background field is large compared to the Compton wavelength of the 
noninteracting fermion, the fermion number is carried intrinsically by
the background field, instead of through explicitly filled levels.

We have found a wide range of model parameters for which we believe that the
soliton is not destroyed by the infrared singularities of the ``would-be''
Goldstone mode, and where the binding of the variational minimum is
sizable.  In this case, the cost in the classical energy is more than
compensated by the gain from the fermion loop.  For a fixed set of the
dimensionless model parameters, the gain in energy is proportional to the
dimensionful Yukawa coupling constant $G$, and the phenomenon of soliton
formation becomes more pronounced as the perturbative mass of the fermion
increases.  Thus, in 1+1 dimensions, heavy fermions can indeed stabilize
solitons.

\subsection*{Acknowledgments} We would like to thank S.~Bashinsky,
J.~Goldstone, and D.~Son for helpful conversations, suggestions, and
references. We thank J.~Feinberg and A.~Zee for helpful correspondence
related to their work.  This work is supported in part by funds provided by
the U.S. Department of Energy (D.O.E.) under cooperative research agreement
\#DF-FC02-94ER40818 and the Deutsche Forschungsgemeinschaft (DFG) under
contract We 1254/3-1.

\appendix

\renewcommand{\theequation}{\thesection.\arabic{equation}}

\section{Dimensional regularization\protect\newline of the Casimir
energy}\label{app:a}

In the body of the paper we introduced and manipulated formally divergent
expressions such as eq.~(\ref{efermion1}) without prescribing a
regularization scheme to render them finite.  For Feynman diagrams,
regularization and renormalization are so familiar that such a casual
approach would not warrant further comment.  In this Appendix we show how
to extend the method of dimensional regularization to Casimir sums, thereby
rendering our earlier manipulations well defined.  This analysis resolves
ambiguities in the definition of the effective energy that have been noted
in recent works \cite{vanN}.

We begin by defining the Casimir energy in $n$ (space) dimensions in terms
of a sum over bound states and an integral over a continuum density of
states.  These are properties of a radial Schr\"odinger-like equation,
which remain well defined when $n$ is noninteger.  We choose $n$ such that
the expressions are finite.  We then isolate and compute the term $\Delta
E^{(1)}$ that diverges when $n$ is continued to $n=1$, the case of
interest.  For $0<n<1$ we show explicitly that $\Delta E^{(1)}$ is
identical to the contribution of the lowest order Feynman diagram in the
loop expansion of the effective energy.  This is our fundamental result. 
Still keeping $0<n<1$ we subtract $\Delta E^{(1)}$ from the Casimir sum and
add back in the Feynman diagram.  The subtracted Casimir sum is finite for
$n=1$ while the now (conventionally) dimensionally regulated Feynman
diagram is renormalized by the counterterm contribution, $\Delta E_{\rm
ct}$.  This calculation demonstrates that we have precisely implemented the
standard dimensional regularization and renormalization process of
perturbative field theory.

For simplicity we consider the self-interactions of a single real 
boson in one dimension; the generalization to fermions is discussed 
in Ref. \cite{tbaglevi}.
We take the bosonic Lagrangian
\begin{equation}
{\cal L} = \half\partial_\mu \phi\, \partial^\mu \phi - U(\phi) + C
U''(\phi)
\label{a1.1}
\end{equation}
where prime denotes differentiation with respect to $\phi$. We have
indicated the counterterm $CU''(\phi)$ explicitly including its
cutoff-dependent coefficient $C$, and we are considering an arbitrary
potential $U(\phi)$.  This counterterm renormalizes the boson's mass, and
is the only counterterm needed to render the theory finite in one spatial
dimension.  We take a background $\phi_0(x)$ that is either a solution to
the classical equations of motion or held in place by an external source,
so that it is a stationary point of the classical action.  We wish to
compute the one-loop contribution to the energy arising from the bosonic
fluctuations about $\phi_0(x)$.  To simplify the analysis, we assume it is
either an odd or an even function of $x$, and we take $U(\phi)$ to be an
even function of $\phi$ that gives a perturbative mass $m$ to $\phi$.
The potential for small oscillations around $\phi_0(x)$ is then given by
\begin{equation}
V(x) = U''(\phi_0(x)) \, .
\end{equation}
Our restrictions on $U(\phi)$ and $\phi_0(x)$ ensure that $V(x)$ is
even in $x$.  As usual, the relationship between the scattering phase
shifts and the density of states is
\begin{equation}
\delta\rho(k) \equiv \rho(k) - \rho_0(k) = \frac{1}{2\pi i}\hbox{Tr}\ln S =
\frac{1}{\pi}\frac{d}{dk} \delta(k) \, .
\label{a1.2}
\end{equation} 
For a single real field $\phi$, we only need to consider positive 
energies.  We use parity to decompose the $S$-matrix into 
symmetric and antisymmetric channels, giving
\begin{equation}
\delta^B(k) = \delta^+(k) + \delta^-(k) \, .
\end{equation}
As in Section III we define the Casimir energy as the sum over
bound states $\{\omega_{j}\}$ plus an integral over the continuum
weighted by $\omega = \sqrt{k^2 + m^2}$,
\begin{equation}
\Delta E[\svec\phi\,]=
\fract{1}{2}\sum_j (\omega_j - m) + \half\int_0^\infty dk
(\omega-m) \delta\rho(k) + \Delta E_{\rm ct} 
\label{cas1B}
\end{equation}
where in analogy to eq.~(\ref{efermion1}) we have used Levinson's 
theorem to regulate potential infrared divergences (as $k\to 0$).

We now generalize eqs.~(\ref{a1.2}) and (\ref{cas1B}) to $n$ dimensions.
The scattering problem generalizes to a central potential in $n$
dimensions.  The $S$-matrix is diagonal in the basis of the irreducible
tensor representations of $SO(n)$. These are the traceless symmetric
tensors of rank $\ell$, where $\ell =  0,1,2,\ldots$.  The formula for the
density of states, eq.~(\ref{a1.2}), becomes a sum over $\ell$,
\begin{equation}
\delta\rho_n(k)
= \frac{1}{\pi} \frac{d}{dk} \delta^B_n(k)
= \frac{1}{\pi} \frac{d}{dk} 
\sum_{\ell=0}^{\infty}N_{n}^{\ell} \delta_{n,\ell}(k)  
\label{a1.3}
\end{equation}
where $\delta_{n,\ell}(k)$ is the phase shift in the $\ell^{\rm th}$ 
partial wave and $N_{n}^{\ell}$ is the degeneracy of the $SO(n)$
representation labelled by $\ell$.  For integer $n$ and $\ell$,
$N_{n}^{\ell}$ is given by the dimension of the space of symmetric tensors
with $\ell$ indices that each run from 1 to $n$, with all traces (contractions)
removed.  Working out the combinatorics gives
\begin{equation}
N_n^\ell = \frac{(n+\ell-1)!}{\ell!(n-1)!} -
\frac{(n+\ell-3)!}{(\ell-2)!(n-1)!} \, .
\label{appb2}
\end{equation}
To prepare the way to continue to noninteger $n$ we write $N_{n}^{\ell}$
in terms of $\Gamma$-functions instead of factorials,
\begin{equation}
N_{n}^{\ell}= 
\frac{\Gamma(n+\ell-2)}{\Gamma(n-1)\Gamma(\ell+1)}(n+2\ell-2) \, .
\label{a1.4}
\end{equation}
For $n=3$, $N_{n}^{\ell}$ reduces to $2\ell+1$ as expected.

The phase shifts are obtained by solving the radial Schr\"odinger
equation generalized to $n$ dimensions,
\begin{equation}
-\psi''-\frac{n-1}{r}\psi' + \frac{\ell(\ell+n-2)}{r^{2}}\psi
+2mV(r)\psi = k^{2}\psi
\label{a1.5}
\end{equation}
which is related to Bessel's equation for $V=0$.  At the origin, the
regular solution $\psi_{n,\ell}$ is proportional to $r^{\ell}$
independent of $n$.

Incoming and outgoing waves are generalizations of spherical Hankel
functions,
\begin{equation}
h^{(1,2)}_{n,\ell}(kr)=\frac{1}{(kr)^{\frac{n}{2}-1}}
\left(J_{\frac{n}{2}+\ell-1}(kr)
\pm i Y_{\frac{n}{2}+\ell-1}(kr)\right)
\label{a1.6}
\end{equation}
and the phase shifts are defined in the usual way by writing the 
solution $\psi_{n,\ell}$ regular at the origin as
\begin{equation} 
	\psi_{n,\ell} \sim h^{(2)}_{n,\ell}(kr)+e^{2i\delta_{n,\ell}(k)}
	h^{(1)}_{n,\ell}(kr) 
	\label{a1.7}
\end{equation}
for large $r$, where the potential vanishes.  The leading behavior of
$\delta_{n,\ell}(k)$ at large $k$ is given by the first Born approximation, 
\begin{equation}
\delta_{n,\ell}^{(1)}(k) = -\frac{\pi}{2}
\int _0^\infty J_{\frac{n}{2} +\ell-1}(kr)^2 V(r) r\, dr \, .
\end{equation}

The expression for the Casimir energy in $n$ dimensions is
\begin{equation}
\Delta E_n[\svec\phi\,]=
\fract{1}{2} \sum_j \Bigl(\sum_{\ell=0}^{\infty} N_n^\ell (\omega_{j,n,\ell}
- m)\Bigr) + \int_0^\infty \frac{dk}{2\pi} (\omega-m)
\sum_{\ell=0}^{\infty} N_n^\ell \frac{d}{dk} \delta_{n,l}(k) +
\Delta E_{{\rm ct},n}
\label{cas1C}
\end{equation}
where $N_n^\ell$ is given by eq.~(\ref{a1.4}), the $\omega_{j,n,\ell}$
are the normalizable solutions to eq.~(\ref{a1.5}) in each partial wave
$\ell$, and $\delta_{n,\ell}(k)$ is determined from eqs.~(\ref{a1.5}) and
eq.~(\ref{a1.7}).  $\Delta E_{{\rm ct},n}$ is the counterterm contribution, to
be fixed by a renormalization condition below.  The rest of eq.~(\ref{cas1C})  is
well defined for noninteger $n$, where the integration over
$k$ and the sum over
$\ell$ converge.  Holding the divergences in abeyance we can verify that
eq.~(\ref{cas1C}) reduces to the naive result, eq.~(\ref{cas1B}), as $n\to
1$:  $N_1^\ell$ vanishes for all $\ell$ except $\ell = 0$ and $\ell = 1$,
where it is $1$; for $\ell = 0$, $\delta_{1,0}(k)$ and $\omega_{j,1,0}$ are
obtained from the solutions to eq.~(\ref{a1.4}) that have vanishing first
derivative at $r=0$, while for $\ell = 1$, $\delta_{1,1}(k)$ and 
$\omega_{j,1,1}$ are obtained from the solutions to eq.~(\ref{a1.4}) that
vanish at $r=0$.  Thus $\ell=0$ and $\ell=1$ correspond to the even and odd
parity channels respectively.

Our approach consists of subtracting the first Born approximation to the
phase shift and replacing its contribution by the tadpole graph, which we
then calculate in ordinary Feynman perturbation theory.  Thus we must
demonstrate explicitly that these two quantities are equal by computing
both as analytic functions of $n$, away from integer $n$, where both
diverge.  Once the leading Born approximation has been subtracted, the
integral over the phase shift is finite and we can take the limit $n\to 1$
with no further subtleties.

The tadpole graph requires the external momentum to be equal to zero. Thus
we should expect that both the leading Born approximation and the tadpole
graph will depend only on the spatial average of the potential,
\begin{equation}
\langle V \rangle = \int V(x)\, d^nx =
\frac{2\pi^\frac{n}{2}}{\Gamma\left(\frac{n}{2}\right)}
\int_0^\infty V(r) r^{n-1}dr \, .
\end{equation}

From eq.~(\ref{cas1C}), the contribution to the energy from the
first Born approximation is
\begin{equation}
\Delta E^{(1)}[\svec\phi\,]=
\sum_{\ell=0}^{\infty} N_\ell^n
\int_0^\infty \frac{dk}{2\pi} (\omega - m)
\frac{d\delta_{n,\ell}^{(1)}(k)}{dk} \, .
\label{fborn}
\end{equation}
We obtained $(\omega - m)$ rather than $\omega$ in the
integrand of eq.~(\ref{fborn}) because of our use of Levinson's theorem to
put the Casimir energy in the form of eq.~(\ref{cas1B}).  This manipulation
was required to ensure the absence of infrared singularities in one spatial
dimension, but we will see that it is also necessary for us to write a sensible
expression in arbitrary dimensions.\footnote{If the boson were {\it
massless}, we would have found an infrared divergence in eq.~(\ref{cas1B})
as $n\to 1$ from the $1/k$ divergence of the Born approximation at $k=0$. 
This divergence reflects the well-known infrared divergences of massless
theories in 1+1~dimensions \cite{Coleman1d}.}

Using the Bessel function identity
\begin{equation}
\sum_{\ell=0}^{\infty}
\frac{(2q+2\ell)\Gamma(2q+\ell)}{\Gamma(\ell+1)}J_{q+\ell}(z)^2 =
\frac{\Gamma(2q+1)}{\Gamma(q+1)^2}
\left(\frac{z}{2}\right)^{2q}
\label{BesselId}
\end{equation}
and setting $q=\frac{n}{2} - 1$, we explicitly sum over $\ell$ in
eq.~(\ref{fborn}) and obtain
\begin{equation}
\Delta E^{(1)}[\svec\phi\,]=
-\frac{\langle V \rangle }{(4\pi)^\frac{n}{2}
\Gamma\left(\frac{n}{2}\right)} (n-2)
\int_0^\infty (\omega-m) k^{n-3}\, dk\, .
\end{equation}
The $k$ integral can be calculated in the vicinity of $n=\fract{1}{2}$ and
then analytically continued, yielding
\begin{equation}
\int_0^\infty (\omega - m) k^{n-3} \, dk = - \frac{m^{n-1}}{4 \sqrt{\pi}}
\Gamma\Bigl(\frac{1-n}{2}\Bigr) \Gamma\Bigl(\frac{n-2}{2}\Bigr)\, .
\end{equation}
Hence we find
\begin{equation}
\Delta E^{(1)}[\svec\phi\,]=
\frac{\langle V \rangle }{(4\pi)^\frac{n+1}{2}}
\Gamma\Bigl(\frac{1-n}{2}\Bigr) m^{n-1}
\end{equation}
which is exactly what we obtain using standard dimensional regularization
for the tadpole diagram in $n+1$ space-time dimensions.

We choose renormalization conditions such that the tadpole graph is exactly
cancelled by the counterterm contribution.  Thus we can implement the
contribution of $\Delta E_{{\rm ct},n}$ by subtracting
$\delta_{n,\ell}^{(1)}(k)$ from $\delta_{n,\ell}(k)$ in eq.~(\ref{cas1C}),
yielding a finite result
\begin{equation}
\Delta E_n[\svec\phi\,]=
\fract{1}{2} \sum_j \sum_{\ell=0}^{\infty} N_n^\ell (\omega_{j,n,\ell} - m)
+ \int_0^\infty \frac{dk}{2\pi} (\omega-m)
\sum_{\ell=0}^{\infty} N_n^\ell
\frac{d}{dk} \left(\delta_{n,l}(k) - \delta^{(1)}_{n,l}(k) \right) \, .
\label{cas1D}
\end{equation}
This result can then be smoothly continued to $n = 1$, giving
\begin{equation}
\Delta E [\svec\phi\,]=
\fract{1}{2}\sum_j (\omega_j - m) + \int_0^\infty \frac{dk}{2\pi}
(\omega-m) \frac{d}{dk} \left(\delta^B(k) - \delta^{(1)}(k) \right)
\label{cas2B}
\end{equation}
where
\begin{equation}
\delta^{(1)}(k) = -\frac{1}{k} \int_0^\infty V(r)\, dr
\, .
\end{equation}

By continuing to fractional dimensions, we have regulated the theory,
rendering it finite.  We made certain that we held the physical
renormalization conditions fixed while removing the regulator.  This process
defines the theory in terms of physical parameters that can be measured
within the perturbative sector.  Based only on these inputs, we can then
calculate the energy of nontrivial field configurations.

\section{Studies of the Tightly Bound Fermion Level}\label{app:b}

Our soliton's stability was driven largely by the strong binding of a
single fermion level in a chiral background.  In this appendix we
explore this phenomenon further by studying the case of a scalar field
constrained to the chiral circle, $\svec\phi = (\cos \Theta(\xi), \sin
\Theta(\xi))$, with the chiral angle $\Theta(\xi)$ varying monotonically
between $0$ and $2\pi$ as $\xi$ goes from $-\infty$ to $+\infty$.  We
assume $\Theta(\xi)=\pi - \Theta(-\xi)$, so $\Theta(0)=\pi$ and
$\Theta'(0)>0$.  Using the WKB approximation, we argue that the vacuum will
acquire unit fermion number when $\Theta(\xi)$ varies slowly compared with $m$.

It is convenient to adopt a basis for the Dirac matrices different
from the body of the paper: $\gamma^0=\sigma_1$, $\gamma_1=i\sigma_2$
and hence $\gamma_5=\sigma_3$.  Denoting the upper and lower
components of the spinor by $a$ and $b$, respectively, the first order
Dirac equations read
\begin{eqnarray}
a&=&e^{-i\Theta}\left(\omega b +i b^\prime\right)\nonumber\\
b&=&e^{i\Theta}\left(\omega a -i a^\prime\right)
\label{forder}
\end{eqnarray} 
where as before, the energy $\omega$ is measured in units of the
fermion mass, $m=G v$ and a prime denotes a derivative with respect to
$\xi=mx$.  In this basis, the second order equations become
\begin{eqnarray}
-a^{\prime\prime}&=&\left(\omega^2-1\right)a
+\Theta^\prime\left(ia^\prime-\omega a\right)
\label{sordera}\\
-b^{\prime\prime}&=&\left(\omega^2-1\right)b
-\Theta^\prime\left(ib^\prime+\omega b\right)\, .
\label{sorderb}
\end{eqnarray}
A bound state is a solution to eq.~(\ref{forder}) which falls exponentially
as $\xi\to\infty$ and has definite parity:  $a(0)=\pm b(0)$.

\begin{figure}[hbt]
\centerline{
\BoxedEPSF{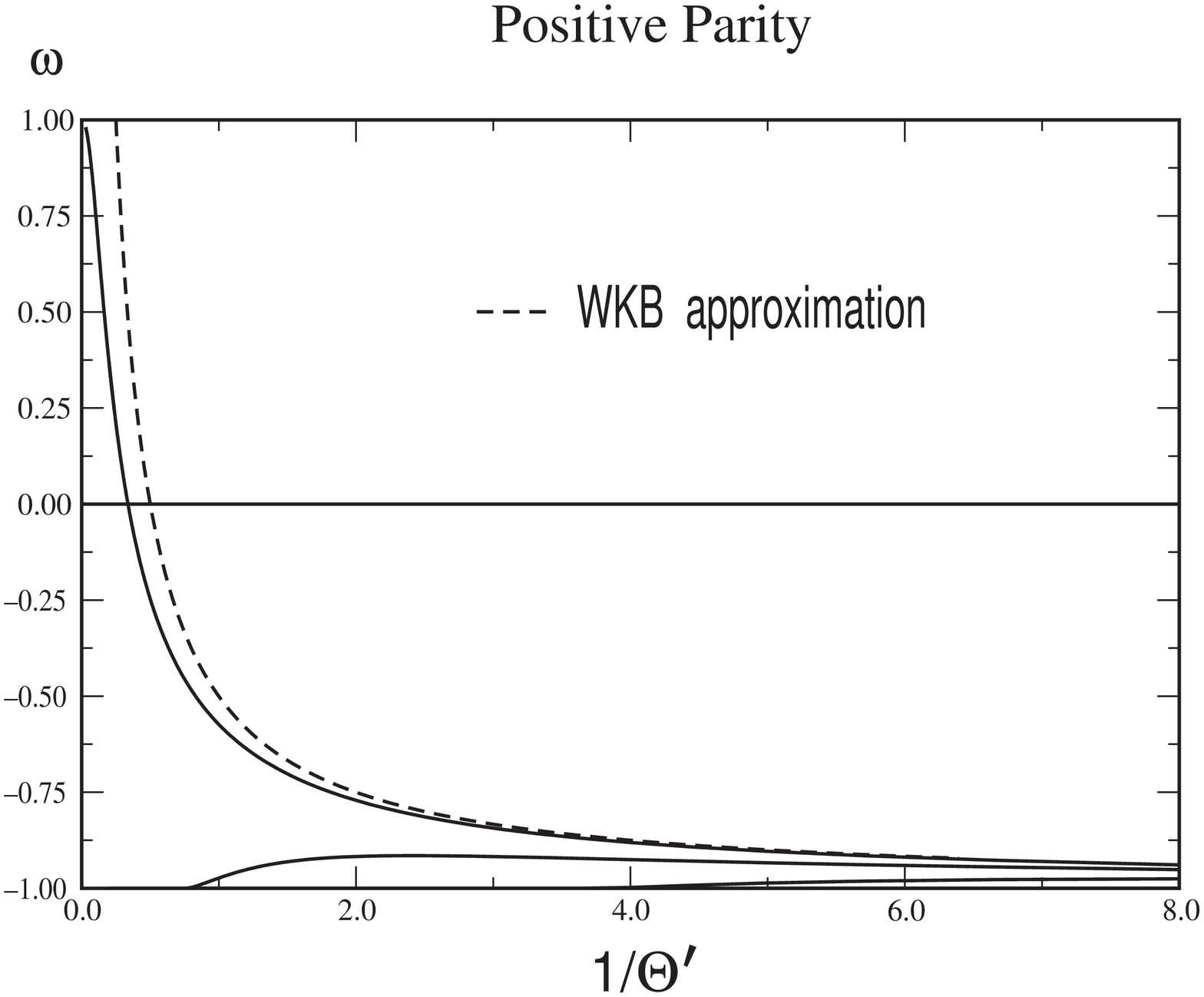 scaled 300}
\quad
\BoxedEPSF{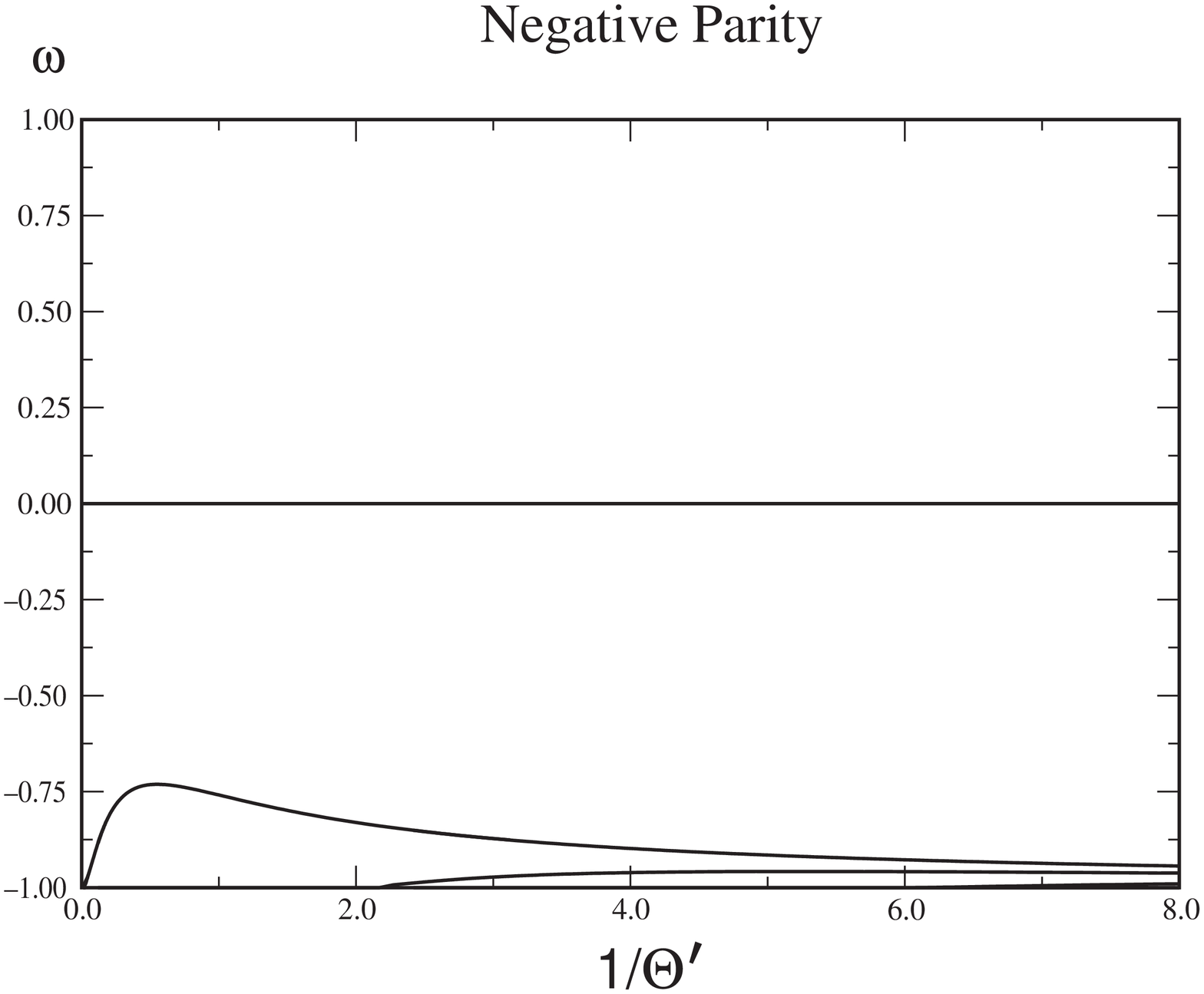 scaled 300}
}\bigskip
\caption{\sl The bound state energies of the simple model as functions of
the inverse of the constant slope $\Theta^\prime$. In the positive parity
channel the WKB solution (\protect\ref{strong}) is also shown.}
\label{toymodel}
\end{figure}

First, we suppose $\Theta$ is slowly varying and use the WKB
approximation.  We parameterize $a(\xi)=e^{if(\xi)}$ and neglect
$f''$ compared with $f^{\prime2}$.  Substituting into eq.~(\ref{sordera})
and solving for $f'$ we obtain
\begin{equation}
f' = -\frac{\Theta'}{2}+i\sqrt{1-\omega^{2}+\omega\Theta'
-\fract{1}{4}\Theta^{\prime2}} 
\label{WKB}
\end{equation}
where we have chosen the root that gives a wavefunction that falls
exponentially at large $\xi$.  $b(\xi)$ is given by the second of
eqs.~(\ref{forder}).  The eigenvalue condition for positive (negative)
parity at $\xi=0$ reduces to
\begin{equation}
-\omega-f'(0) = \pm 1\ .
\label{WKBeigen}
\end{equation}
This equation has only one solution when $\Theta'(0)>0$,
\begin{equation}
\omega = -1 +\Theta'(0)/2  
\label{strong}
\end{equation}
which has positive parity.  $1/\Theta'(0)$ measures the width of the
scalar field configuration.  For large width, the WKB solution lies just
above the negative energy continuum.  As the width decreases, the WKB bound
state energy increases, crosses zero when $\Theta'(0)=2$ and enters the
positive energy continuum when $\Theta'(0)=4$.

The condition that $f''<f^{\prime2}$ limits the validity 
of the WKB approximation to $\Theta''\ll 1$.  For example, when 
$\omega=0$ the condition reduces to 
\begin{equation}
\left|\frac{\Theta''(\xi)}{\sqrt{4-\Theta^{\prime2}(\xi)}}\right|\ll1\ .
\label{validity}
\end{equation}
This condition can be satisfied for all $\xi$ by making $\Theta''$ 
small enough.  Note $\Theta''(0)=0$ by symmetry thereby avoiding the 
apparent singularity at $\xi=0$ where $\omega=0$ requires 
$\Theta'(0)=2$.

We have augmented the WKB analysis by solving a simple toy model where 
much of the calculation can be done analytically.  In this model we  let
$\Theta'=\gamma=$constant for $-\pi/\gamma<\xi<\pi/\gamma$ and  zero
elsewhere.  The positive and negative parity eigenstates are determined by
a simple transcendental equation which can be solved numerically.  The
results are shown along with the WKB estimate, eq.~(\ref{strong}), in 
Fig.~\ref{toymodel}.  As expected, the toy model has a positive  parity
bound state that descends rapidly from the positive energy  threshold,
through zero energy, toward the negative energy continuum.   It is well
approximated by the WKB estimate.  The model has  other positive and
negative energy bound states, which are missed by  the WKB approximation,
but which remain in the vicinity of either the  positive or negative energy
continuum for all values of $\gamma$.

\section{Reflection Coefficients}\label{app:c}

\begin{figure}[hbt]
\centerline{
\BoxedEPSF{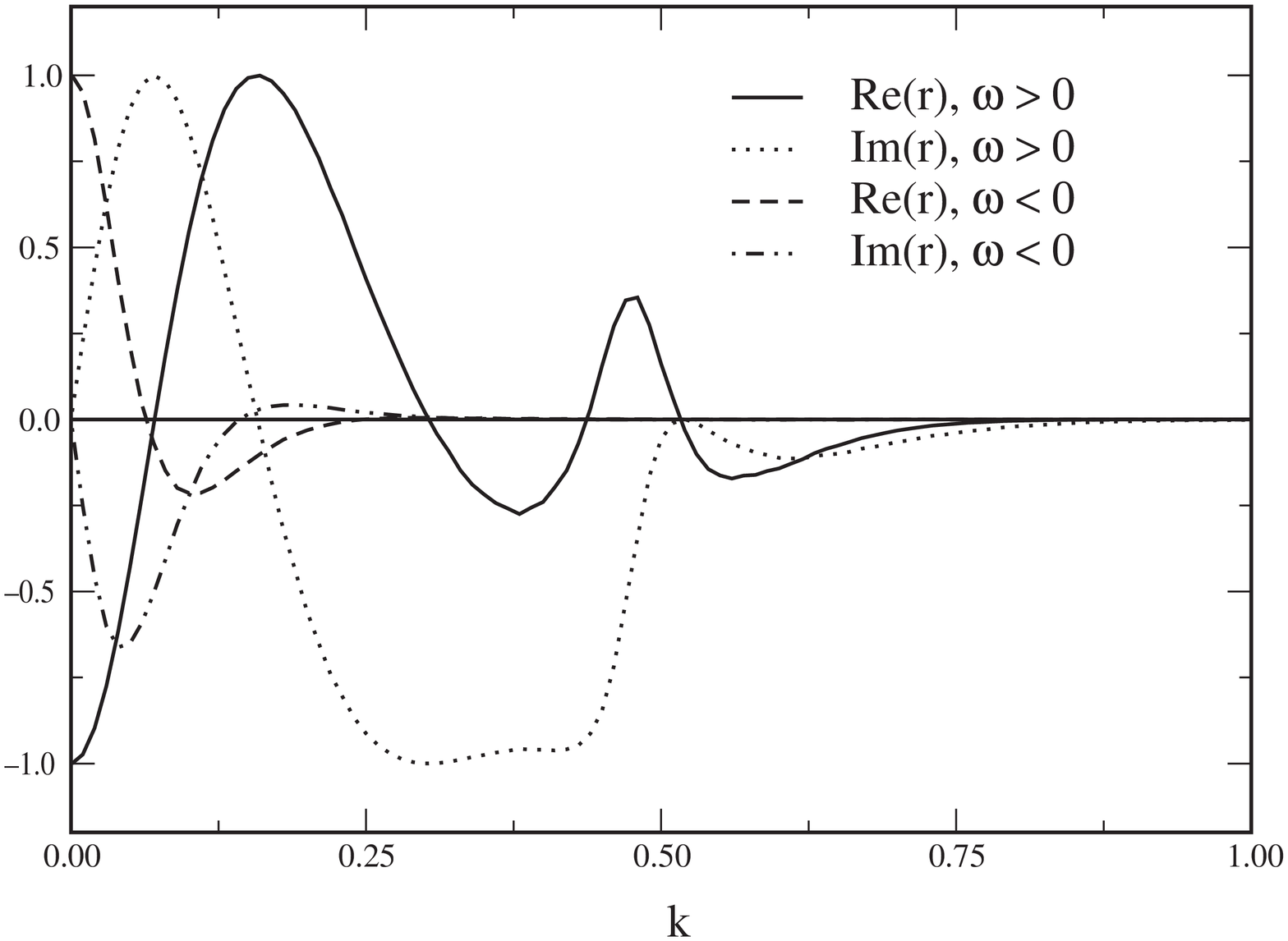 scaled 300}
\quad
\BoxedEPSF{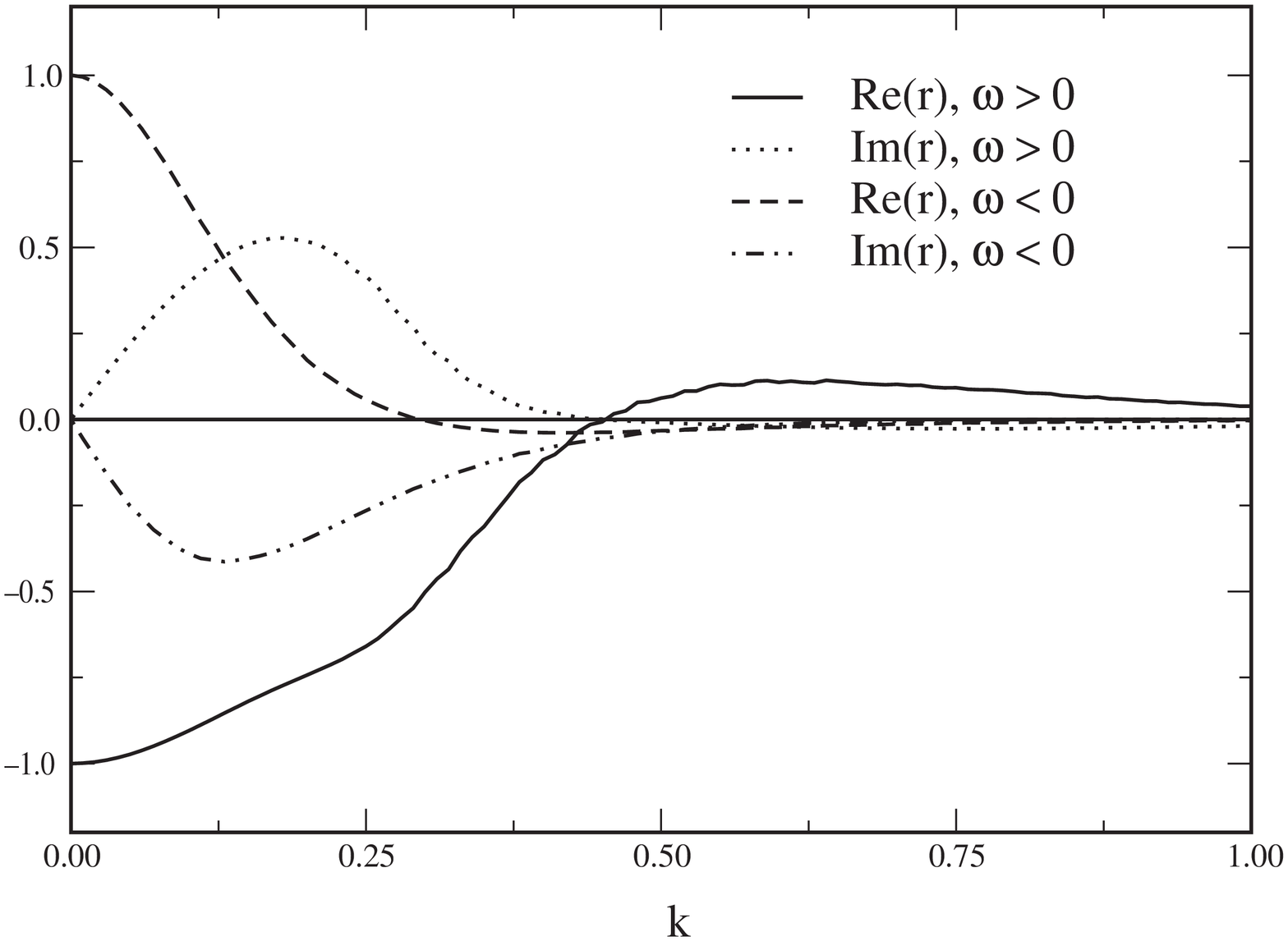 scaled 300}
}\bigskip
\caption{\sl The reflection coefficients of the variational approximation
to the soliton as functions of the momentum $k=\sqrt{\omega^2-1}$. We
consider two values of $\tilde{\alpha}$: $0.1$ (left panel) and $0.6$
(right panel). In both cases we have used $\tilde{\lambda}=1.0$ and
$v/\sqrt{N_F}=0.375$, as in Fig.~\ref{soliton}.}
\label{reflectioncoeff}
\end{figure}

Ref.~\cite{Zee} claimed that the reflection coefficient for fermions
scattering off a soliton background should vanish.  Since we calculate the
phase shifts $\delta_\pm(\omega)$, we can easily compute the reflection
coefficient
\begin{equation}
r(\omega)=\fract{1}{2}\left[S_+(\omega)-S_-(\omega)\right]
\label{defrefl}
\end{equation}
in terms of the S-matrix elements $S_\pm(\omega) =
\exp(2i\delta_\pm(\omega))$. We are thus equipped to reexamine this
statement for our variational approximation to the soliton. In
Fig.~\ref{reflectioncoeff} we display the complex reflection coefficients
as functions of the momentum of the scattering fermion for positive and
negative energies. Although these coefficients approach zero very quickly
as $k$ increases, they do not vanish identically. In
particular we find that these coefficients acquire more structure as we get
closer to the (unstable) symmetric formulation ($\tilde{\alpha}=0$). The
added structure is mainly caused by the increasing number of bound states
in the wider potential.  Because of Levinson's theorem these bound
states cause the reflection coefficient to circle around the origin in the
complex plane.  Though it is still possible that the true soliton gives a
reflectionless potential, we see no indication of this behavior.

The statement in Ref.~\cite{Zee} is based on the correct requirement that
the fermion current
\begin{equation}
j^1(x) = (N_F-{\cal Q})\psi_1(x)^\dagger \gamma^0 \gamma^1 \psi_1(x)
+\langle \Omega | \Psi(x)^\dagger \gamma^0 \gamma^1 \Psi(x) |
\Omega \rangle
\end{equation}
must vanish for all $x$ in the soliton background.  However,
decomposing $\Psi(x)$ in terms of creation and annihilation  operators that
multiply the solutions to the Dirac equation given in eq.~(\ref{sdef}), we
find that $j^1(x)$ vanishes mode by mode as a simple consequence of $|S_+|
= |S_-| = 1$ which is enforced by unitarity. Thus the requirement
$j^1(x)\equiv 0$ is trivially fulfilled and places no restriction on the
soliton configuration.

Though two of the most famous examples of topological solitons, the kink
and the sine-Gordon soliton, both correspond to reflectionless potentials, 
we have no reason to assume that such a restriction applies in general.  
Indeed, the Jacobi model studied in Ref.~\cite{Dunne} gives a family
of solitons labelled by the elliptic parameter $\nu$.  These solitons
correspond to potentials that are not reflectionless except in the limit
$\nu\to 0$ and $\nu\to 1$ (where the model reduces to the sine-Gordon
case).

For certain  interactions, the inverse scattering  method
applies and the equations of motion (expressed in terms of scattering data)
indeed yields reflectionless potentials \cite{DHN}.  However, the attempts
to extend this method to  chiral models in
Ref.~\cite{Shei,Campbell} are problematic:  First, they cannot include the
symmetry breaking terms needed to avoid the restoration of chiral symmetry
by boson loops.  Second, because these solitons approach different
configurations as $x\to-\infty$ and $x\to+\infty$ (which are degenerate
vacua in the absence of symmetry breaking), the methods developed in
Ref.~\cite{tbaglevi,GoldWil} show that the polarized Dirac sea carries
fractional charge for each fermion flavor even though no level has crossed
zero.  When this contribution is properly included, the total fermion
number of the configuration constructed in Ref.~\cite{Shei} adds to zero.


\begin{thebibliography}{99}
\frenchspacing 

\bibitem{method1}
E.~Farhi, N.~Graham, P.~Haagensen and R.L.~Jaffe,
Phys.~Lett.~{\bf B427} (1998) 334.

\bibitem{method2}
N.~Graham and R.L.~Jaffe, Nucl. Phys. {\bf B544} (1999) 432.

\bibitem{method3}
N.~Graham and R.L.~Jaffe, Nucl. Phys. {\bf B549} (1999) 516.

\bibitem{tbaglevi}
E. Farhi, N. Graham, R.L. Jaffe and H. Weigel,
``Fractional and Integer Charges from Levinson's Theorem'',
in preparation.

\bibitem{prevwork}
R.~MacKenzie, F.~Wilczek, and A.~Zee, Phys.~Rev.~Lett.~{\bf 53} (1984) 2203.
G.~Ripka and S.~Kahana, Phys.~Lett.~{\bf 155B} (1985) 327,
Phys.~Rev.~{\bf D36} (1987) 1233;
R.~J.~Perry, Nucl.~Phys.~{\bf A467} (1987) 717;
B.~Moussallam, Phys.~Rev.~{\bf D40} (1989) 3430;
G.~Anderson, L.~Hall, and S.~G.~Hsu, Phys.~Lett.~{\bf 249B} (1990) 505;
F.~Wilczek, IAS preprint, IASSNS-HEP-90/20.
S.~Dimopoulos, B.~Lynn, S.~Selipsky, and N.~Tetradis, 
Phys.~Lett.~{\bf 253B} (1991) 237;
J.~Bagger and S.~Naculich, Phys.~Rev.~Lett.~{\bf 67}
(1991) 2252, Phys.~Rev.~{\bf D45} (1992) 1395.
S.~G.~Naculich, Phys.~Rev~{\bf D46} (1992) 5487.

\bibitem{Farhi} E.~D'Hoker and E.~Farhi, Nucl.~Phys.~{\bf B248} (1984) 59,
E.~D'Hoker and E.~Farhi, Nucl.~Phys.~{\bf B248} (1984) 77.

\bibitem{vanN}
A.~Rebhan and P.~van Nieuwenhuizen, Nucl.~Phys.~{\bf B508} (1997) 449;
H.~Nastase, M.~Stephanov, P.~van Nieuwenhuizen and A.~Rebhan, 
Nucl.~Phys.~{\bf B542} (1999) 471.

\bibitem{Shei}
S.~Shei, Phys.~Rev.~{\bf D14} (1976) 535.

\bibitem{Zee}
J.~Feinberg and A.~Zee, Int.~J.~Mod.~Phys.~{\bf A12} (1997) 1133;
J.~Feinberg and A.~Zee, Phys.~Rev.~{\bf D56} (1997) 5050.

\bibitem{Campbell}
D.K.~Campell and Yao-Tang~Liao, Phys.~Rev.~{\bf D14} (1976) 2093.

\bibitem{letter}
E.~Farhi, N.~Graham, R.L.~Jaffe and H.~Weigel, 
Phys.~Lett.~{\bf B475} (2000) 335. 

\bibitem{Coleman1d}
S.~Coleman, Commun.~Math.~Phys.~{\bf 31} (1973) 259.

\bibitem{Herbert}
R.~Alkofer, H.~Reinhardt, and H.~Weigel, Phys.~Rept.~{\bf 265} (1996) 139.

\bibitem{selfcoupled}
N.~Graham and R.~L.~Jaffe, Phys.~Lett.~{\bf B435} (1998) 145.

\bibitem{Cole}  S.~Coleman,
 {\sl Aspects of Symmetry} (Cambridge University Press, Cambridge, 1985);
(North-Holland, Amsterdam, 1982).

\bibitem{Barton}
G. Barton, J.~Phys.~A: Math.~Gen.~{\bf 18} (1985) 479.

\bibitem{GoldWil}
J.~Goldstone and F.~Wilczek, Phys.~Rev.~Lett.~{\bf 47} (1981) 986.

\bibitem{JR}
R.~Jackiw and C.~Rebbi, Phys.~Rev.~{\bf D13} (1976) 3398.

\bibitem{Dunne} G.V.~Dunne, Phys.~Lett.~{\bf B467} (1999) 238.

\bibitem{DHN} R.F.~Dashen, R.~Hasslacher, and A.~Neveu,
Phys.~Rev.~{\bf D12} (1975) 2443.

\end{thebibliography}
\end{document}